\documentclass[twocolumn]{autart}    % Enable this line and disable the 
\usepackage[utf8]{inputenc}   % This two packages are used for \v and \'
\usepackage[T1]{fontenc}      % same as previous

\usepackage{amsmath, rotating}
\usepackage{array, multirow}
\usepackage{amsbsy, amsfonts, graphics}
\usepackage{algorithmic}
\usepackage{algorithm}
\usepackage{epstopdf}
\usepackage{mathrsfs}
\usepackage{graphicx}
\usepackage{caption,subcaption}
\usepackage{amssymb}
\usepackage{amsmath,bm}
\usepackage{subfiles}
\usepackage{mathtools}

\usepackage{tikz}
\usepackage{float}
\usetikzlibrary{calc,patterns,decorations.pathmorphing,decorations.markings}
\usepackage{soul}
\usepackage{cancel}
\usepackage{todonotes}

\usepackage{subfig}
\usepackage[normalem]{ulem} % 给删除线准备的，haing [normalem] makes \emph{} to be italic not underline.
\usepackage{units}
\newcommand{\bea}{\begin{eqnarray}}
\newcommand{\eea}{\end{eqnarray}}
\newcommand{\beas}{\begin{eqnarray*}}
	\newcommand{\eeas}{\end{eqnarray*}}

\definecolor{mygreen}{RGB}{0,180,0}

\usepackage{wrapfig}
\usepackage{graphicx}                              % document contains figures,
\makeatletter
\renewcommand{\maketag@@@}[1]{\hbox{\m@th\normalsize\normalfont#1}}%
\makeatother

\begin{document}

\begin{frontmatter}

\title{Model-Free Output Tracking of Nonlinear Systems via High-Frequency Dither\thanksref{footnoteinfo}} % Title, preferably not more 

\thanks[footnoteinfo]{This work was supported by the Australian Research
Council under the Discovery Project DP260101027.}

\author[Melbourne]{Weixuan Wang}\ead{weixuanw1@unimelb.edu.au},
\author[Melbourne]{Ying Tan}\ead{yingt@unimelb.edu.au},
\author[Melbourne]{Ye Pu}\ead{ye.pu@unimelb.edu.au}

%\address[Melbourne]{School of Electrical, Mechanical and Infrastructure Engineering, The University of Melbourne, Parkville, 3010, Victoria, Australia}
\address[Melbourne]{Faculty of Engineering and Information Technology, The University of Melbourne, Parkville, 3010, Victoria, Australia}

% \address[Paestum]{Buckingham Palace, Paestum}  % Please supply                                              
% \address[Rome]{Senate House, Rome}             % full addresses
% \address[Baiae]{The White House, Baiae}        % here.

                             % keyword list or with the 
                                          % help of the Automatica 
                                          % keyword wizard

\begin{abstract}                      
This paper investigates model-free output tracking of time-varying reference trajectories for a class of unknown nonlinear control-affine systems with known relative degree one. The tracking problem is formulated as the minimisation of a time-varying objective function, and a high-frequency-dither-based controller is developed using only measured output performance, without requiring knowledge of the plant dynamics or control direction. By deliberately introducing high-frequency excitation, the proposed approach establishes an exploitable separation of time scales. A coordinate transformation inspired by vibrational control removes the explicit high-frequency input term, enabling partial averaging of the nonlinear closed-loop dynamics. For analysis, the partial averaging step reveals an effective gradient-descent-like tracking mechanism whose stabilising action is independent of the sign of the unknown input–output gain. Under suitable assumptions on the internal dynamics, semiglobal practical exponential convergence guarantees on the tracking of time-varying reference trajectories is established for appropriately selected controller parameters and sufficiently high dither frequencies. While sharing the optimisation-based perspective of extremum-seeking control, the proposed approach addresses time-varying optimisation in the presence of unknown nonlinear plant dynamics without relying on an input to steady-state output mapping. By exploiting transient plant dynamics through high-frequency excitation, the proposed framework provides a foundation for extending the approach to more general time-varying performance objectives and incorporating input and output constraints.

\end{abstract}

\end{frontmatter}

% \section{Introduction}
% Video, patres conscripti, in me omnium vestrum ora atque oculos esse 
% conversos, video vos non solunn de vestro ac rei publicae, verum 
% etiam, si id depulsum sit, de meo periculo esse sollicitos. Est mihi 
% iucunda in malis et grata in dolore vestra erga me voluntas, sed eam, 
% per deos inmortales, deponite atque obliti salutis meae de vobis ac 
% de vestris liberis cogitate. Mihi si haec condicio consulatus data 
% est, ut omnis acerbitates, onunis dolores cruciatusque perferrem, 
% feram non solum fortiter, verum etiam lubenter, dum modo meis 
% laboribus vobis populoque Romano dignitas salusque pariatur.

% \begin{figure}
% \begin{center}
% \includegraphics[height=4cm]{jcaesar.eps}    % The printed column  
% \caption{Gaius Julius Caesar, 100--44 B.C.}  % width is 8.4 cm.
% \label{fig1}                                 % Size the figures 
% \end{center}                                 % accordingly.
% \end{figure}

% OR

%\begin{figure}
%\begin{center}
%\epsfig{file=jcaesar,width=7cm}
%\caption{Gaius Julius Caesar, 100--44 B.C.}
%\label{fig1}
%\end{center}
%\end{figure}

\section{Introduction}
High-frequency dither offers an additional degree of freedom beyond conventional state and output feedback for shaping the stability and dynamic behaviour of nonlinear systems \cite{meerkov1977vibrational,meerkov1980principle,bellman1986stabilizability,cheng2018robustness}. Rapid excitation can induce motion that alters how physical forces act on a system, thereby creating control effects that are not available through the nominal feedback action alone. Even small-amplitude vibrations can produce significant dynamic effects at sufficiently high frequencies, depending on how the excitation interacts with the underlying system dynamics. This principle has been explored in nonlinear control, including the smoothing of discontinuous nonlinearities in relay feedback systems \cite{zames1976dither,iannelli2003dither}, compensation of nonlinear actuator characteristics such as friction and hysteresis, and vibrational stabilization of dynamical systems \cite{bellman1986stabilizability,cheng2022robustness}. These developments demonstrate the potential of high-frequency excitation as a complementary mechanism for control design.

Beyond shaping nonlinear system dynamics, high-frequency excitation has been used for full-state trajectory tracking with unknown control directions \cite{scheinker2012extremum} and for output regulation of exosystem-generated signals using an internal model \cite{wang2024}. Output tracking is fundamental to applications such as robotic motion control, where accurate tracking is required despite modelling uncertainties and disturbances \cite{slotine1983tracking,slotine1989composite}. When the plant dynamics are only partially known, sliding-mode control can stabilise the tracking-error dynamics \cite{shtessel2014sliding,levant2003higher}, while funnel control and prescribed-performance control enforce bounds on the tracking error \cite{berger2021funnel,zhang2024tracking}. A time-varying optimisation formulation provides a way to address tracking performance, control effort, and operational constraints within the same framework.

From this optimisation perspective, we address model-free tracking of time-varying references for unknown relative-degree-one nonlinear systems. The controller uses scalar output measurements and requires no model of the reference generator. We consider bounded, absolutely continuous references with essentially bounded derivatives, including signals that may vary on time scales shorter than those of the open-loop plant dynamics.
%Motivated by this optimisation perspective, this paper investigates model-free output tracking of time-varying references for unknown nonlinear control-affine systems with relative degree one. 
Tracking is formulated as the minimisation of a time-varying performance function, and a high-frequency-dither-based controller is developed using only measured output performance, without requiring a plant model or knowledge of the control direction. The deliberately introduced high-frequency dither establishes a separation of time scales between the rapid oscillatory excitation and the slower evolution of the transformed closed-loop dynamics and time-varying reference trajectory. This separation motivates the use of partial averaging \cite{Teel2003Unified}, which characterises the effects of fast oscillations while retaining the evolution of the remaining dynamical variables and the time dependence of the reference trajectory. The resulting averaged dynamics reveal an effective gradient-descent-like tracking mechanism whose stabilising action is independent of the sign of the unknown input--output gain.

% \red{The reference is assumed to have bounded amplitude and a bounded rate of change, with no requirement that the latter bound be small. Consequently, rapidly varying references can be accommodated through appropriate selection of the controller gain and dither frequency. }

The main contribution is a partial-averaging-based framework that reveals an effective gradient-descent-like mechanism for model-free tracking of time-varying references in unknown nonlinear systems. A coordinate transformation inspired by vibrational control removes the explicit high-frequency input term, enabling partial averaging of the nonlinear closed-loop dynamics \cite{meerkov1980principle,bellman1986stabilizability}. The resulting analysis establishes semiglobal practical exponential tracking under suitable assumptions on the internal dynamics, for appropriately selected controller parameters and sufficiently high dither frequencies. Unlike classical extremum-seeking control (ESC) \cite{krstic2000stability,tan2006non}, the proposed approach exploits transient nonlinear plant dynamics without requiring a quasi-steady-state input--output map, complementing recent developments in fast ESC \cite{guay2017proportional,liu2024novel}. The potential of high-frequency excitation to stabilise otherwise unstable internal dynamics is illustrated through a numerical example in Section~\ref{subsec:vibrational_stabilisation}, while the optimisation-based formulation provides a foundation for incorporating more general performance objectives and input and output constraints.

\textbf{Notation:}
Let $L_\infty$ be the set of all essentially bounded measurable functions. For any $\bm{x}(\cdot) \in L_{\infty} $, $|\bm{x}|_\infty\coloneqq\operatorname*{ess\,sup}_{t\geq0}|\bm{x}(t)|$ denotes the $L_\infty$ norm.
%
%For a function $f(t):\mathbb{R}_{\geq 0} \rightarrow \mathbb{R}_{\geq 0}$, $D^+ f$ denotes the upper right Dini derivative of $f$.

\section{Problem Setting} \label{Section Problem setting}

Consider the unknown single-input single-output nonlinear control-affine system
\begin{equation}
    \label{eq:plant}
    \begin{aligned}
        \dot{\bm{x}} &= \bm{f}(\bm{x})+\bm{g}(\bm{x})u,\\
        y &= h(\bm{x}),
    \end{aligned}
\end{equation}
where $\bm{x}\in\mathbb{R}^n$ is the state, $u\in\mathbb{R}$ is the control input, and $y\in\mathbb{R}$ is the measured output.
The vector fields $\bm{f},\bm{g}:\mathbb{R}^n\to\mathbb{R}^n$ are of class $C^2$, and the output mapping $h:\mathbb{R}^n\to\mathbb{R}$ is of class $C^3$.
Only the output $y$ is available for feedback, and the mappings $\bm{f}$, $\bm{g}$, and $h$ are unknown.

The unknown plant \eqref{eq:plant} is assumed to satisfy the following condition.

\begin{assum}
    \label{assum:Relative_degree}
    The system \eqref{eq:plant} has relative degree one on $\mathbb{R}^n$: 
    \begin{equation}
        \label{eq:Lgh}
        L_gh(\bm{x})
        =
        \frac{\partial h}{\partial\bm{x}}(\bm{x})\bm{g}(\bm{x})
        \neq 0,
        \qquad \forall\,\bm{x}\in\mathbb{R}^n,
    \end{equation}
    where $\bm{g}(\cdot)$ and $h(\cdot)$ come from (\ref{eq:plant}).
    \hfill$\triangle$
\end{assum}

\begin{rem}
    Relative degree one arises naturally in systems where
    the control input directly influences the rate of change
    of the measured output, such as thermal processes,
    chemical reactors, and mechanical systems with velocity
    outputs.
    This assumption is standard in classical funnel
    control and first-order sliding-mode control
    \cite{berger2021funnel,slotine1983tracking},
    where the sign of the input--output gain
    $L_gh(\bm{x})$ is typically assumed to be known.
    Both approaches have been extended to higher relative
    degrees \cite{berger2021funnel,levant2003higher}.
    In contrast, the proposed controller does not require
    knowledge of the control direction, providing a
    possible direction for extending the framework
    to higher-relative-degree systems.
\end{rem}

Under Assumption~\ref{assum:Relative_degree},
the well-known Byrnes--Isidori coordinate transformation
exists locally \cite[Theorem 13.1]{khalil2002nonlinear}.
Specifically, on a neighbourhood
$\mathcal U\subseteq\mathbb R^n$, there exists
a diffeomorphism
\begin{equation}
    \bm{\Theta}(\bm{x})
    \coloneqq
    \begin{bmatrix}
        \bm{\vartheta}(\bm{x})\\
        h(\bm{x})
    \end{bmatrix},
    \qquad
    \bm{\vartheta}:\mathcal U\to\mathbb R^{n-1},
    \label{eq:BI_transformation}
\end{equation}
where
\begin{equation}
    \frac{\partial\bm{\vartheta}}{\partial\bm{x}}
    (\bm{x})\bm{g}(\bm{x})=\bm{0}.
    \label{eq:BI_eta_condition_x}
\end{equation}

Defining the new coordinates
\begin{equation}
    \begin{bmatrix}
        \bm{\eta}\\
        y
    \end{bmatrix}
    =
    \bm{\Theta}(\bm{x}),
    \qquad
    \bm{x}=\bm{\Theta}^{-1}(\bm{\eta},y),
    \label{eq:BI_averaged_coordinates}
\end{equation}
the plant \eqref{eq:plant} can be expressed locally
in Byrnes--Isidori normal form as
\begin{equation}
    \begin{aligned}
        \dot{\bm{\eta}} &= \bm{q}(\bm{\eta},y),\\
        \dot y &= \ell(\bm{\eta},y)+b(\bm{\eta},y)u,
    \end{aligned}
    \label{eq:BI_system}
\end{equation}
where
\begin{align}
    \bm{q}(\bm{\eta},y)
    &\coloneqq
    \left.
    \frac{\partial\bm{\vartheta}}{\partial\bm{x}}
    (\bm{x})\bm{f}(\bm{x})
    \right|_{\bm{x}=\bm{\Theta}^{-1}(\bm{\eta},y)},
    \label{eq:q_definition}\\
    \ell(\bm{\eta},y)
    &\coloneqq
    L_fh\bigl(\bm{\Theta}^{-1}(\bm{\eta},y)\bigr),
    \label{eq:ell_definition}\\
    b(\bm{\eta},y)
    &\coloneqq
    L_gh\bigl(\bm{\Theta}^{-1}(\bm{\eta},y)\bigr),
    \label{eq:b_definition}
\end{align}
and $b(\bm \eta,y) \neq 0$ under Assumption~\ref{assum:Relative_degree}.

The following assumption strengthens the local
coordinate property and characterises the boundedness
of the internal dynamics.

\begin{assum}   \label{assum:Internal_dynamics}
    The transformation $\bm{\Theta}$ in
    \eqref{eq:BI_transformation} is a global
    diffeomorphism from $\mathbb{R}^n$ onto
    $\mathbb{R}^n$.
    Moreover, there exist a continuously differentiable,
    positive definite, radially unbounded function
    $W:\mathbb{R}^{n-1}\to\mathbb{R}_{\geq0}$,
    a class-$\mathcal K_\infty$ function $\alpha$,
    a continuous nondecreasing function
    $\sigma:\mathbb{R}_{\geq0}\to\mathbb{R}_{\geq0}$,
    and a constant $c_0\geq0$ such that
    \begin{equation}
        \label{eq:internal_boundedness}
        \frac{\partial W}{\partial\bm{\eta}}
        (\bm{\eta})\bm{q}(\bm{\eta},y)
        \leq
        -\alpha(W(\bm{\eta}))
        +\sigma(|y|)+c_0
    \end{equation}
    for all $(\bm{\eta},y)\in
    \mathbb{R}^{n-1}\times\mathbb{R}$.
    \hfill$\triangle$
\end{assum}

\begin{rem}
Assumption~\ref{assum:Internal_dynamics} imposes two
restrictive conditions. First, the global Byrnes--Isidori
transformation, satisfied by the chemical reactor in
Section~\ref{subsec:illustrative_example}, may be relaxed to a local transformation
if the closed-loop trajectories remain within its domain,
with corresponding restrictions on the tracking guarantee.

Second, bounded output does not necessarily imply bounded
internal states. The Lyapunov condition
\eqref{eq:internal_boundedness} ensures their boundedness
for bounded outputs, as in funnel control
\cite{berger2021funnel}. This condition may potentially
be relaxed through vibrational stabilisation, as
illustrated in Section~\ref{subsec:vibrational_stabilisation}.
\end{rem}

The desired output trajectory is also assumed to satisfy the following condition.

\begin{assum}
    \label{assum:Reference_trajectory}
    The reference trajectory $y_d:[0,\infty)\to\mathbb{R}$ is absolutely continuous, with constants $R,D\geq 0$ such that
    \begin{equation}
        \label{eq:reference_bounds}
        |y_d|_\infty\leq R,
        \qquad
        |\dot y_d|_\infty\leq D.
    \end{equation}
    \hfill$\triangle$
\end{assum}

\begin{rem}
    Assumption \ref{assum:Reference_trajectory} is
    satisfied by a broad class of reference trajectories
    arising in practical applications, including periodic
    signals and bounded-speed motion profiles.
    Such regularity conditions are standard in nonlinear
    tracking control; see, e.g.,
    \cite{berger2021funnel}.
    No higher-order derivatives of the reference are required.
\end{rem}
With the plant and reference trajectory specified,
the output-tracking problem is formulated as follows.

\noindent\textbf{Control objective.}
Define the output tracking error as
\begin{equation}
    e(t)\coloneqq y(t)-y_d(t).
    \label{eq:tracking_error}
\end{equation}
Under Assumptions~\ref{assum:Relative_degree}--\ref{assum:Reference_trajectory}, the objective is to
achieve semiglobal practical exponential output tracking
using only the measured output and reference trajectory,
without knowledge of the plant dynamics. Specifically,
for any prescribed compact set of initial conditions
and any $\rho,\delta>0$, the closed-loop system should
satisfy
\begin{equation}
    |e(t)|\leq
    e^{-\rho(t-t_0)}|e(t_0)|+\delta,
    \qquad t\geq t_0,
    \label{eq:tracking_design_objective}
\end{equation}
while ensuring boundedness of all closed-loop states.
Here, $\rho$ denotes the prescribed exponential decay
rate and $\delta$ the ultimate tracking tolerance.

\section{Main Result}

This section presents a high-frequency-dither-based controller for model-free output tracking of time-varying references. The deliberately introduced dither enables time-scale separation, which is exploited through a coordinate transformation and partial averaging to reveal an effective gradient-descent-like tracking mechanism. Semiglobal practical exponential tracking is established under suitable assumptions on the internal dynamics.

\subsection{Controller Design} \label{subsec:controller_design}

We formulate the output-tracking problem as the minimisation of the time-varying objective function
\begin{equation}
    J(t,y)=J(t,h(\bm{x}))\coloneqq\bigl(y-y_d(t)\bigr)^2,
    \label{eq:tracking_objective}
\end{equation}
where the time-varying reference $y_d(t)$ defines the moving minimiser of $J(t,y)$. Although the mappings $\bm{f}$, $\bm{g}$, and $h$ are unknown and the state $\bm{x}$ is not measured, the objective function can be evaluated directly from the measured output $y$ and the reference trajectory $y_d(t)$.

\begin{figure}[!htb]
    \centering
    \includegraphics[width=\linewidth]{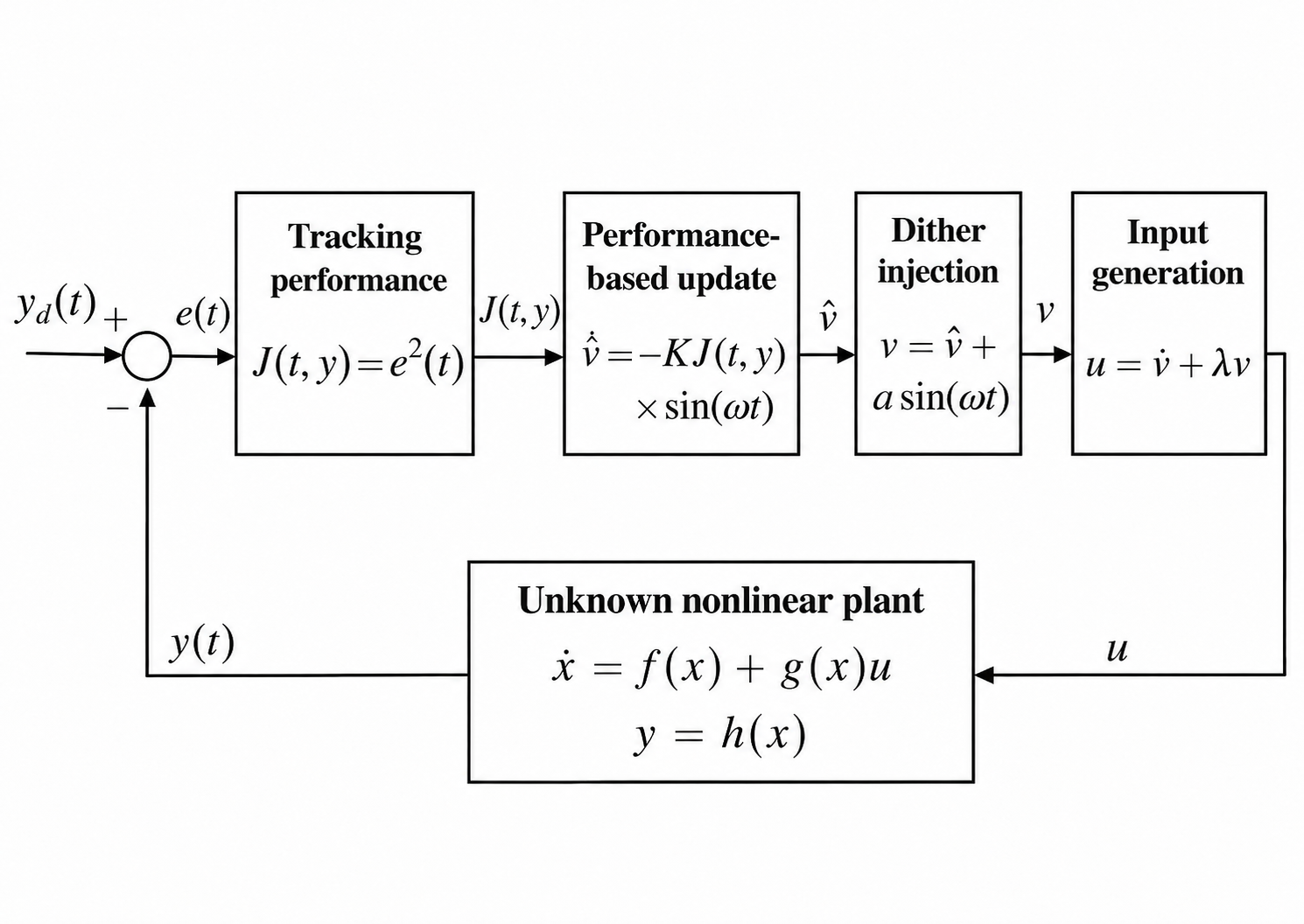}
    \caption{Proposed model-free output-tracking
    control architecture.}
    \label{fig:block_diagram}
\end{figure}

We propose the high-frequency-dither-based
output-tracking controller illustrated in
Fig.~\ref{fig:block_diagram}:
\begin{equation}
    \label{eq:control_law}
    \begin{aligned}
        \dot{\hat v} &=-KJ(t,y)\sin(\omega t),\\
        v&=\hat v+a\sin(\omega t),\\
        u&=\dot v+ \lambda v,
    \end{aligned}
\end{equation}
where $a>0$ and $\omega>0$ denote the dither amplitude
and frequency, respectively, and $K>0$ and $\lambda>0$
are controller gains. These are user-selected design
parameters, whose selection influences the closed-loop
performance. 
For simplicity of analysis, we set
$\lambda=1$ throughout the remainder of the paper.
Although the choice of $\lambda$ may affect the
transient response, it is not expected to alter
the qualitative stability properties of the
closed-loop system.
By \eqref{eq:control_law}, the input can be evaluated as
\bea
        u
        =-KJ(t,y)\sin(\omega t)+a\omega\cos(\omega t)+\hat v+ a\sin(\omega t).
        \label{eq:control_input_explicit}
\eea
Substituting \eqref{eq:control_input_explicit} into
the plant \eqref{eq:plant}, the resulting closed-loop
system is given by
\begin{equation}
    \label{eq:closed_loop_system}
    \begin{aligned}
        \dot{\bm{x}}
        ={}&\bm{f}(\bm{x})
        +\bm{g}(\bm{x})
        \bigl[
            -KJ(t,h(\bm{x}))\sin(\omega t)\\
        &\quad
            +a\omega\cos(\omega t)
            +\hat v
            + a\sin(\omega t)
        \bigr],\\
        \dot{\hat v}
        ={}&-KJ(t,h(\bm{x}))\sin(\omega t).
    \end{aligned}
\end{equation}
The closed-loop system \eqref{eq:closed_loop_system}
contains the high-frequency term
$a\omega\cos(\omega t)$, whose amplitude grows
with $\omega$.
Motivated by the coordinate transformations used in
vibrational control \cite{bellman1986stabilizability},
we introduce a change of coordinates to eliminate
this explicit high-frequency term.
\begin{rem}
High-frequency dither has long been used in nonlinear control to smooth relay nonlinearities, compensate actuator nonlinearities, and achieve vibrational stabilisation \cite{zames1976dither,iannelli2003dither,meerkov1980principle,bellman1986stabilizability}. Coordinate transformations for removing explicit high-frequency excitation are also well established in vibrational control \cite{meerkov1980principle,bellman1986stabilizability}. In this work, these established techniques are employed to facilitate partial averaging of the nonlinear closed-loop dynamics.
\end{rem}

Consider the flow $\bm{\varphi}(s,\bm{\xi})$
generated by the input vector field $\bm{g}$,
defined by
\begin{equation}
    \label{eq:flow_definition}
    \frac{\partial\bm{\varphi}}{\partial s}
    (s,\bm{\xi})
    =
    \bm{g}\bigl(\bm{\varphi}(s,\bm{\xi})\bigr),
    \qquad
    \bm{\varphi}(0,\bm{\xi})=\bm{\xi}.
\end{equation}
Let
\begin{equation}
    s(t)=a\sin(\omega t),
    \label{eq:dither_signal}
\end{equation}
and introduce the time-varying coordinate transformation
\begin{equation}
    \label{eq:flow_transformation}
    \bm{x}
    =
    \bm{\varphi}\bigl(s(t),\bm{\xi}\bigr).
\end{equation}
The coordinate transformation \eqref{eq:flow_transformation} is locally invertible on any prescribed compact set for sufficiently small dither amplitude $a$, by standard properties of flows of smooth vector fields \cite[Sec.~6.2, Thm.~6.1]{teschl2012ordinary}. Following the coordinate transformation approach commonly employed in vibrational control \cite{meerkov1980principle,bellman1986stabilizability}, it eliminates the explicit high-frequency term $a\omega\cos(\omega t)$ from \eqref{eq:closed_loop_system}, yielding the transformed closed-loop system:
\bea
               \dot{\bm{\xi}}
        &=&
        \bm f_1(s,\bm\xi)
        +
        \bm{g}(\bm{\xi})
        \bigl(\dot{\hat v}+\hat v+s\bigr), \nonumber\\
                \dot{\hat v}
        &=&
        -KJ\bigl(t,h(\bm{\varphi}(s,\bm{\xi}))\bigr)
        \sin(\omega t). \label{eq:transformed_system}
\eea
where 
\begin{equation}
    \bm f_1(s,\bm\xi)
    \coloneqq
    \bigl[
        \nabla_{\bm\xi}\bm\varphi(s,\bm\xi)
    \bigr]^{-1}
    \bm f\bigl(\bm\varphi(s,\bm\xi)\bigr).
    \label{eq:f1_definition}
\end{equation}
Importantly, the coordinate transformation removes the explicit high-frequency term whose amplitude grows with $\omega$, but retains the fast oscillatory dependence through the transformed dynamics, including the output mapping $y=h(\bm{\varphi}(s,\bm{\xi}))$. The effects of the remaining oscillations can then be characterised through partial averaging with respect to the fast phase $\omega t$. The detailed derivation is provided in Appendix~\ref{Proof of Theorem}.

\subsection{Tracking Guarantee}

Define the effective gain
\begin{equation}
\kappa\coloneqq\frac{Ka}{2}.
\end{equation}
The following theorem establishes semiglobal practical exponential tracking for the proposed closed-loop system, with the controller gain $K$ determined by the selected values of $\kappa$ and $a$.

\begin{thm}
\label{thm:tracking_SPES_transfer}
Suppose Assumptions~\ref{assum:Relative_degree}--\ref{assum:Reference_trajectory} hold.
For any prescribed $\Delta,\rho,\delta>0$,
there exist constants $\kappa^*>0$ and $a^*>0$
such that, for every $\kappa\geq\kappa^*$ and
$0<a\leq a^*$, there exists $\omega^*>0$
such that, for every $\omega\geq\omega^*$,
all solutions of the closed-loop system
\eqref{eq:closed_loop_system} satisfying
\[
    \left|
    \begin{bmatrix}
        \bm{x}(t_0)\\
        \hat v(t_0)
    \end{bmatrix}
    \right|\leq\Delta
\]
are forward complete, with all closed-loop states
remaining bounded, and the tracking error satisfying
\begin{equation}
    |e(t)|
    \leq e^{-\rho(t-t_0)}|e(t_0)|+\delta,
        \label{eq:original_tracking_transfer_bound}
\end{equation}
for all $t\geq t_0$. \hfill$\square$
\end{thm}
\noindent\textbf{Proof:} The proof of Theorem~\ref{thm:tracking_SPES_transfer} is given in Appendix~\ref{Proof of Theorem}.

\begin{rem}
Theorem~\ref{thm:tracking_SPES_transfer} requires
sequential selection of the controller parameters:
for each admissible choice of $\kappa$ and $a$,
a sufficiently large dither frequency $\omega$
must be selected. In particular, the threshold
$\omega^*$ need not be uniform in $\kappa$ and $a$,
and may require adjustment when these parameters
change \cite{tan2006non}.
\end{rem}
\begin{rem}
Unlike classical ESC \cite{krstic2000stability,tan2006non},
the proposed approach exploits transient nonlinear plant
dynamics without relying on a quasi-steady-state
input--output map. While fast ESC methods relax this
requirement, they primarily address time-invariant
objectives \cite{moase2012fast,guay2017proportional}.
In contrast to funnel control \cite{berger2021funnel},
which enforces prescribed tracking-error bounds,
the proposed optimisation-based formulation offers
a potential framework for incorporating more general
objectives and constraints \cite{tan2013extremum,ames2017cbf}.
Rigorous guarantees for these extensions remain
to be established.
\end{rem}
%
\iffalse
\begin{rem}
    The high-frequency dither introduces a fast time
    scale $\tau=\omega t$, over which the time-varying
    objective function can be regarded as approximately
    time-invariant, owing to the bounded rate of change
    of the reference trajectory.
    Following the vibrational coordinate transformation,
    averaging reveals an effective tracking feedback
    mechanism involving the square of the unknown
    input--output gain.
    This mechanism drives the output towards the
    instantaneous minimiser of the objective function,
    without requiring knowledge of the plant dynamics
    or control direction, or a steady-state
    input--output map.
\end{rem}
\fi
\begin{rem}\label{rem:Vibrational_control}
High-frequency excitation may stabilise otherwise
unstable internal dynamics, potentially relaxing
Assumption~\ref{assum:Internal_dynamics}.
Such a relaxation requires stability analysis of
the averaged internal dynamics and corresponding
tracking guarantees. This possibility is illustrated
numerically in Section~\ref{subsec:vibrational_stabilisation}.
\end{rem}

\begin{rem}\label{rem:time-varying cost}
The proposed dither-based mechanism may extend to
general time-varying optimisation without plant dynamics,
using only objective-value measurements. Establishing
tracking guarantees requires suitable conditions on
the gradient towards the moving minimiser
\cite{tan2006non}, together with regularity and
boundedness assumptions. This possibility is
illustrated numerically in Section~\ref{subsec:time_varying_objective}.
\end{rem}

\section{Illustrative Examples}
\label{Section Illustrative Example}

Three numerical examples are presented to illustrate
the proposed approach. The first example demonstrates
practical output tracking of a time-varying reference
for a nonlinear chemical reactor, examining the
effects of controller parameters and the sign of
the input--output gain. The second example illustrates
the possibility of relaxing
Assumption~\ref{assum:Internal_dynamics}
through vibrational stabilisation, while maintaining
output tracking. The third example demonstrates
the applicability of the proposed control scheme
in Fig.~\ref{fig:block_diagram} to the optimisation
of a general time-varying cost function.

\subsection{Illustration of Theorem~\ref{thm:tracking_SPES_transfer}}
\label{subsec:illustrative_example}

Consider a stirred-tank reactor adapted from \cite{Yu2012}:
\begin{equation}
    \begin{aligned}
        \dot{x}_1 &= -k(T)x_1+d(x_{1,\mathrm{in}}-x_1),\\
        \dot{x}_2 &= k(T)x_1-dx_2,\\
        \dot{T} &= \beta k(T)x_1-q_TT+u,\qquad y=T,
    \end{aligned}
    \label{eq:cstr_plant}
\end{equation}
where $x_1$ and $x_2$ denote the reactant and product concentrations, respectively, in $\mathrm{mol/L}$, $T$ is the reactor temperature in $\mathrm{K}$, and time is measured in seconds. The unconstrained thermal input $u\in\mathbb{R}$ is measured in $\mathrm{K/s}$. The plant parameters are
\[
    x_{1,\mathrm{in}}=1,\quad
    d=\frac{1.1}{60},\quad
    q_T=\frac{1.25}{60},\quad
    \beta=209.2,
\]
with the temperature-dependent reaction rate
\[
    k(T)=\frac{\exp(25-8700/T)}{60},\qquad T>0,
\]
smoothly extended by $k(T)=0$ for $T\leq0$.
The plant satisfies Assumptions~\ref{assum:Relative_degree} and \ref{assum:Internal_dynamics}, with $L_gh=1$ and global normal-form coordinates $\bm{\eta}=(x_1,x_2)$ and $y=T$.

We select the time-varying reference trajectory
$y_d(t)=300+5\sin(2t)$, which satisfies
Assumption~\ref{assum:Reference_trajectory} with
$R=305$ and $D=10$.
The initial conditions are
$(x_1(0),x_2(0),T(0))=(0.02,0.9,303)$
and $\hat v(0)=0$.

\begin{figure}[!htb]
    \centering
    \begin{subfigure}[b]{0.49\linewidth}
        \centering
        \includegraphics[width=\linewidth]{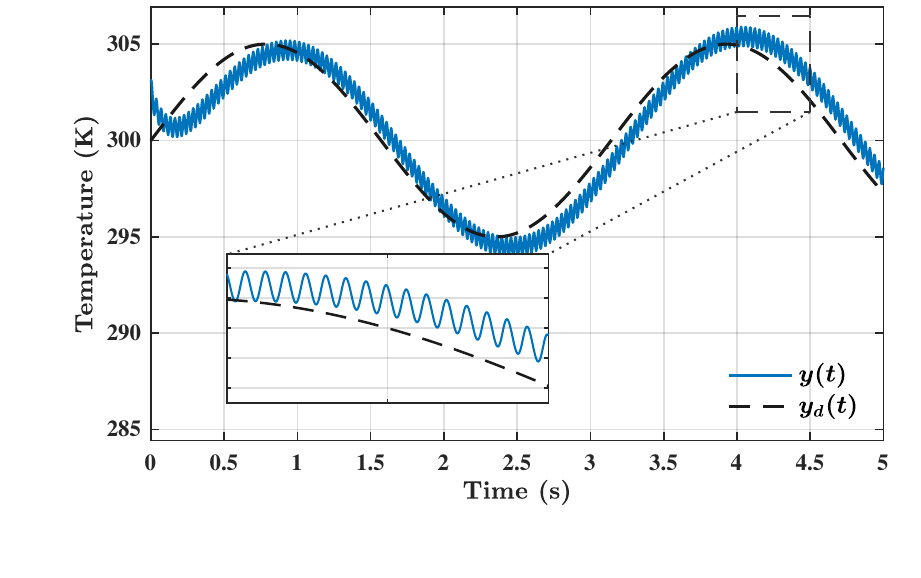}
        \caption{Tracking performance}
        \label{fig:nominal_tracking}
    \end{subfigure}\hfill
    \begin{subfigure}[b]{0.49\linewidth}
        \centering
        \includegraphics[width=\linewidth]{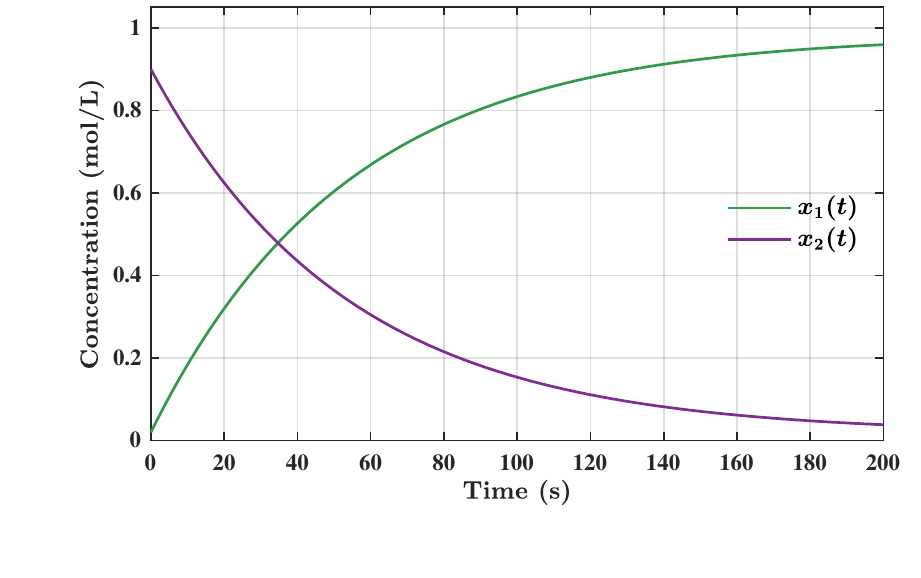}
        \caption{Internal states}
        \label{fig:nominal_internal_states}
    \end{subfigure}
    \caption{Control performance under nominal parameter values.}
\end{figure}

Since Assumptions~\ref{assum:Relative_degree}--\ref{assum:Reference_trajectory} hold for this example, Theorem~\ref{thm:tracking_SPES_transfer} guarantees semiglobal practical exponential tracking for appropriately selected controller parameters. We implement the proposed control scheme in Fig.~\ref{fig:block_diagram} with the nominal parameters $a=0.5$, $K=16$, and $\omega=200\,\mathrm{rad/s}$, corresponding to $\kappa=Ka/2=4$. Figure~\ref{fig:nominal_tracking} shows the resulting output tracking over $0\leq t\leq 5\,\mathrm{s}$, while Fig.~\ref{fig:nominal_internal_states} shows the internal-state trajectories over $0\leq t\leq200\,\mathrm{s}$, which remain bounded over the simulation interval.

\begin{figure}[!htb]
    \centering
    \begin{subfigure}[b]{0.49\linewidth}
        \centering
        \includegraphics[width=\linewidth]{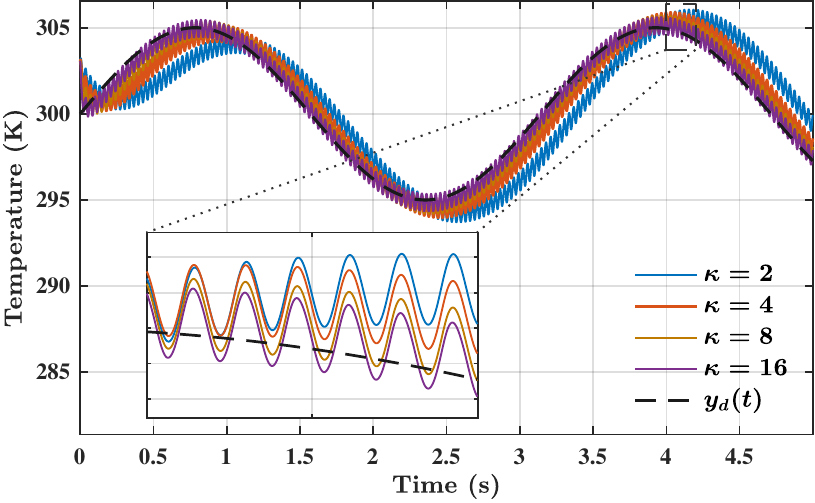}
        \caption{Output tracking for varying $\kappa$.}
        \label{fig:CTRS_kappa_y}
    \end{subfigure}\hfill
    \begin{subfigure}[b]{0.49\linewidth}
        \centering
        \includegraphics[width=\linewidth]{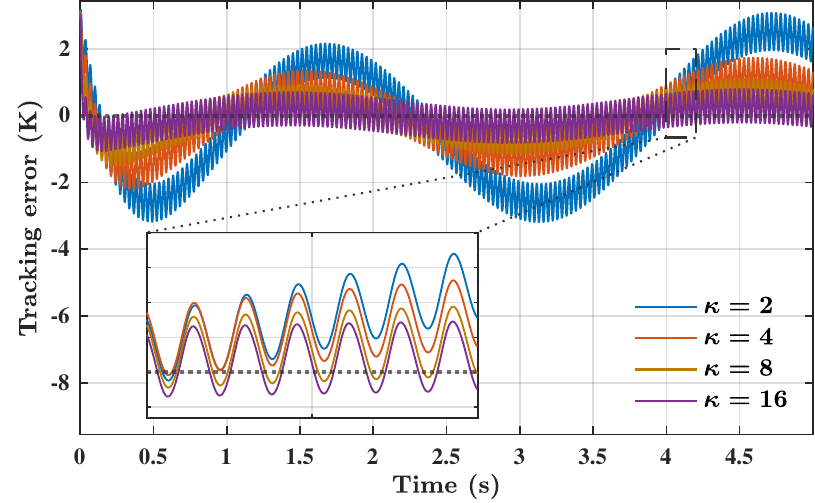}
        \caption{Tracking error for varying $\kappa$.}
        \label{fig:CTRS_kappa_e}
    \end{subfigure}

    \par\medskip

    \begin{subfigure}[b]{0.49\linewidth}
        \centering
        \includegraphics[width=\linewidth]{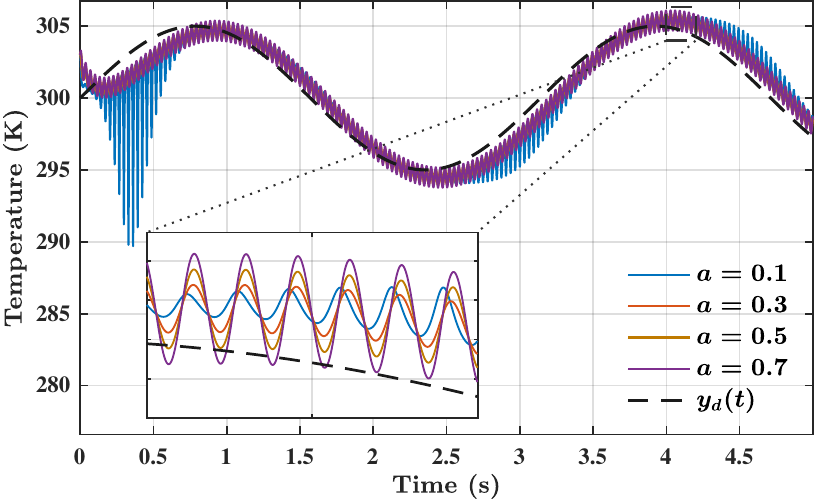}
        \caption{Output tracking for varying $a$.}
        \label{fig:CTRS_a_y}
    \end{subfigure}\hfill
    \begin{subfigure}[b]{0.49\linewidth}
        \centering
        \includegraphics[width=\linewidth]{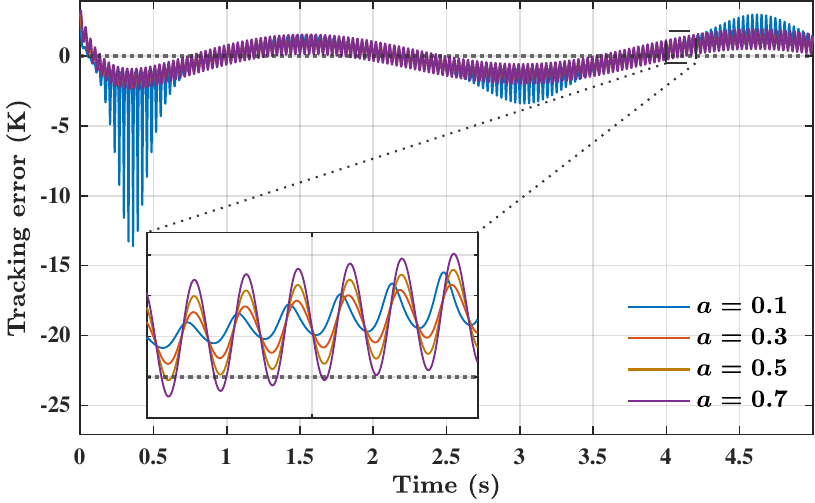}
        \caption{Tracking error for varying $a$.}
        \label{fig:CTRS_a_e}
    \end{subfigure}

    \par\medskip

    \begin{subfigure}[b]{0.49\linewidth}
        \centering
        \includegraphics[width=\linewidth]{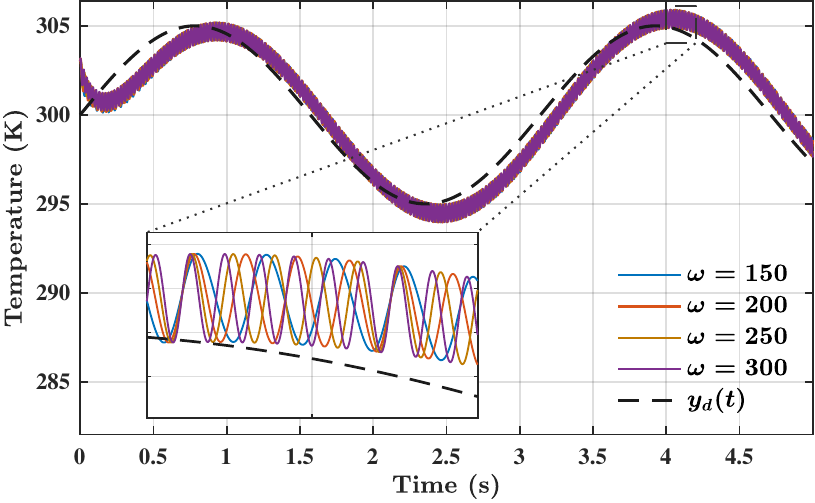}
        \caption{Output tracking for varying $\omega$.}
        \label{fig:CTRS_omega_y}
    \end{subfigure}\hfill
    \begin{subfigure}[b]{0.49\linewidth}
        \centering
        \includegraphics[width=\linewidth]{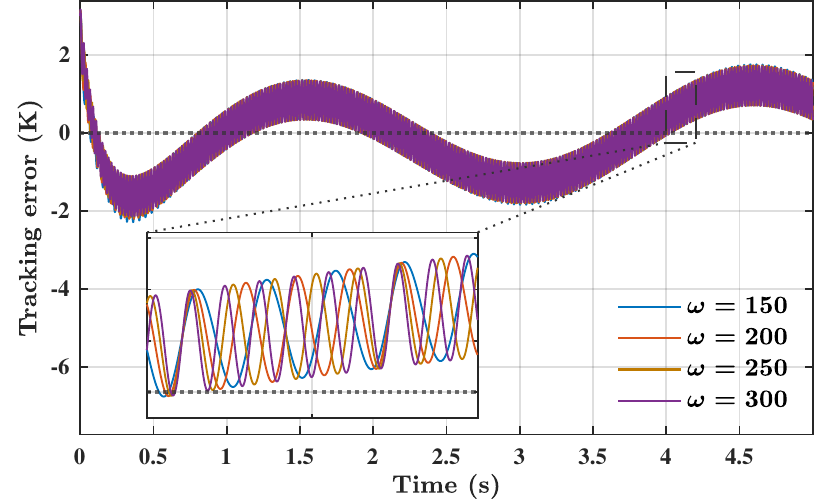}
        \caption{Tracking error for varying $\omega$.}
        \label{fig:CTRS_omega_e}
    \end{subfigure}

    \caption{Effects of controller parameter tuning on
    the closed-loop tracking performance of the nonlinear
    reactor. The left column shows the output $y(t)$
    and reference $y_d(t)$, while the right column shows
    the tracking error $e(t)$. The three rows illustrate
    the effects of varying $\kappa$, $a$, and $\omega$,
    respectively, with the remaining parameters held
    at their nominal values. 
The magnified views highlight differences in the
tracking responses.
}
    \label{fig:CTRS}
\end{figure}

\begin{figure}[t]
    \centering
    \includegraphics[width=0.8\linewidth]{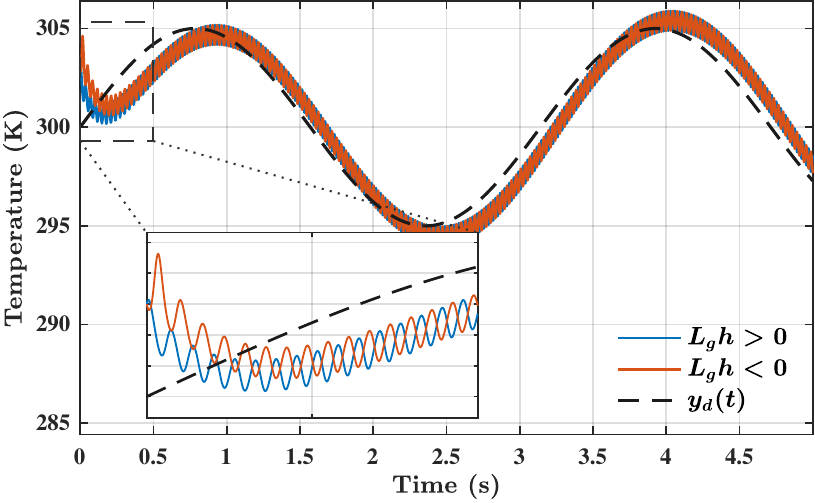}
    \caption{Temperature tracking under different
    input--output gain signs.}
    \label{fig:CTRS_Lgh}
\end{figure}

To illustrate the controller tuning guidelines established in Theorem~\ref{thm:tracking_SPES_transfer}, we investigate the effects of $\kappa$, $a$, and $\omega$ on tracking performance in Fig.~\ref{fig:CTRS}. We further examine the tracking performance under opposite signs of the input--output gain in Fig.~\ref{fig:CTRS_Lgh}.

In the parameter studies, one of the independent parameters $\kappa$, $a$, and $\omega$ is varied at a time, while the other two remain at their nominal values; the gain $K=2\kappa/a$ is adjusted accordingly. Figures~\ref{fig:CTRS_kappa_y} and~\ref{fig:CTRS_kappa_e} show that increasing $\kappa$ reduces the tracking error after the initial transient. This observation is consistent with the damping term in the averaged error dynamics~\eqref{eq:averaged_BI_error}, whose coefficient is proportional to $\kappa$. Increasing $a$ reduces the initial oscillations but increases the amplitude of the high-frequency oscillations during subsequent tracking, as shown in Figs.~\ref{fig:CTRS_a_y} and~\ref{fig:CTRS_a_e}. For the tested parameter values, increasing $\omega$ at fixed $\kappa$ and $a$ also reduces the initial oscillations and slightly improves subsequent tracking, as shown in Figs.~\ref{fig:CTRS_omega_y} and~\ref{fig:CTRS_omega_e}.

To investigate the effect of the input--output gain sign, we compare the original reactor with $L_gh=1$ and a modified reactor obtained by replacing $+u$ with $-u$ in~\eqref{eq:cstr_plant}, giving $L_gh=-1$. The same control law and controller parameters are applied in both cases, without requiring knowledge of the control direction. Figure~\ref{fig:CTRS_Lgh} shows comparable practical tracking performance for both signs of the input--output gain, with modest differences in the initial transients. These results illustrate the ability of the proposed controller to accommodate either control direction without modifying its structure or retuning its parameters.

\begin{figure}[!htb]
    \centering
    \begin{subfigure}[b]{0.49\linewidth}
        \centering
        \includegraphics[width=\linewidth]{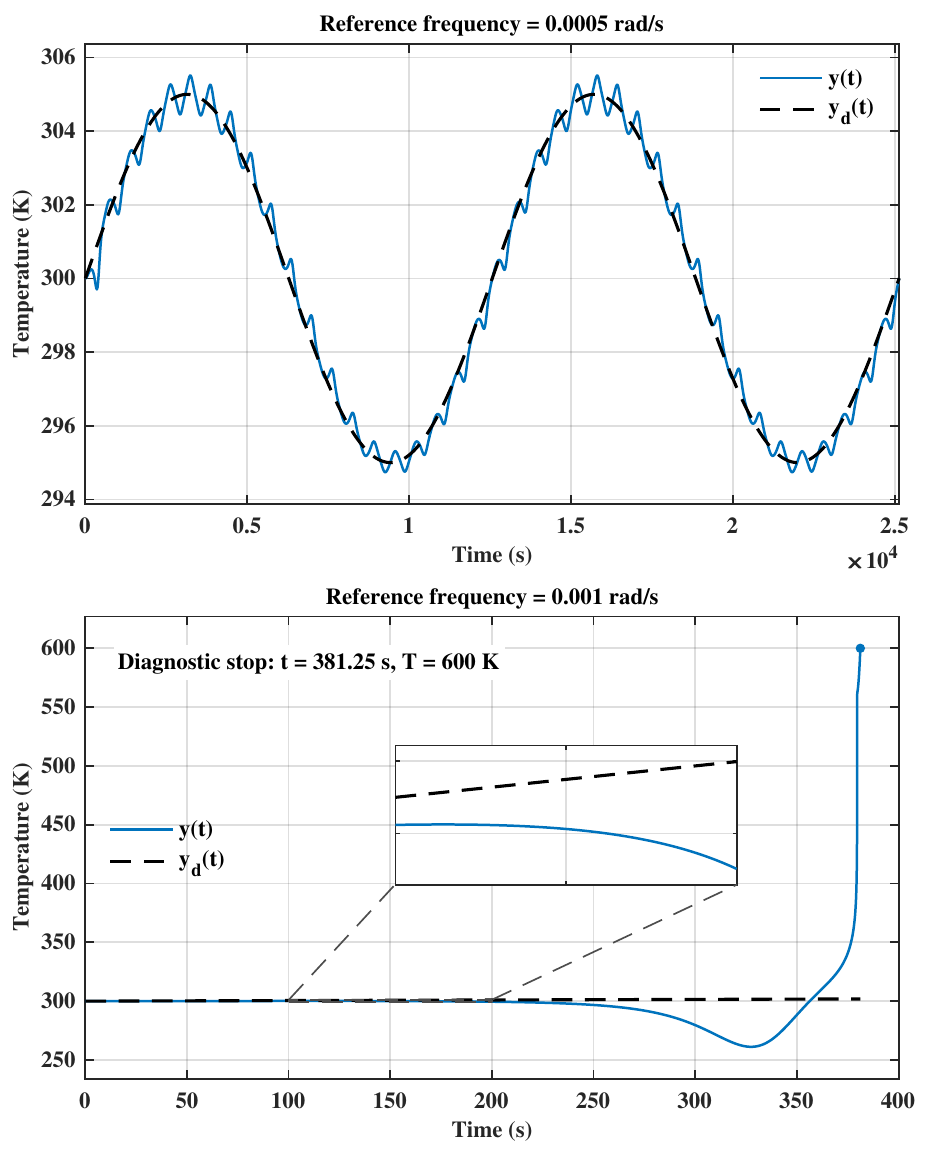}
        \caption{Classical ESC.}
        \label{fig:classical_ESC_baseline}
    \end{subfigure}\hfill
    \begin{subfigure}[b]{0.49\linewidth}
        \centering
        \includegraphics[width=\linewidth]{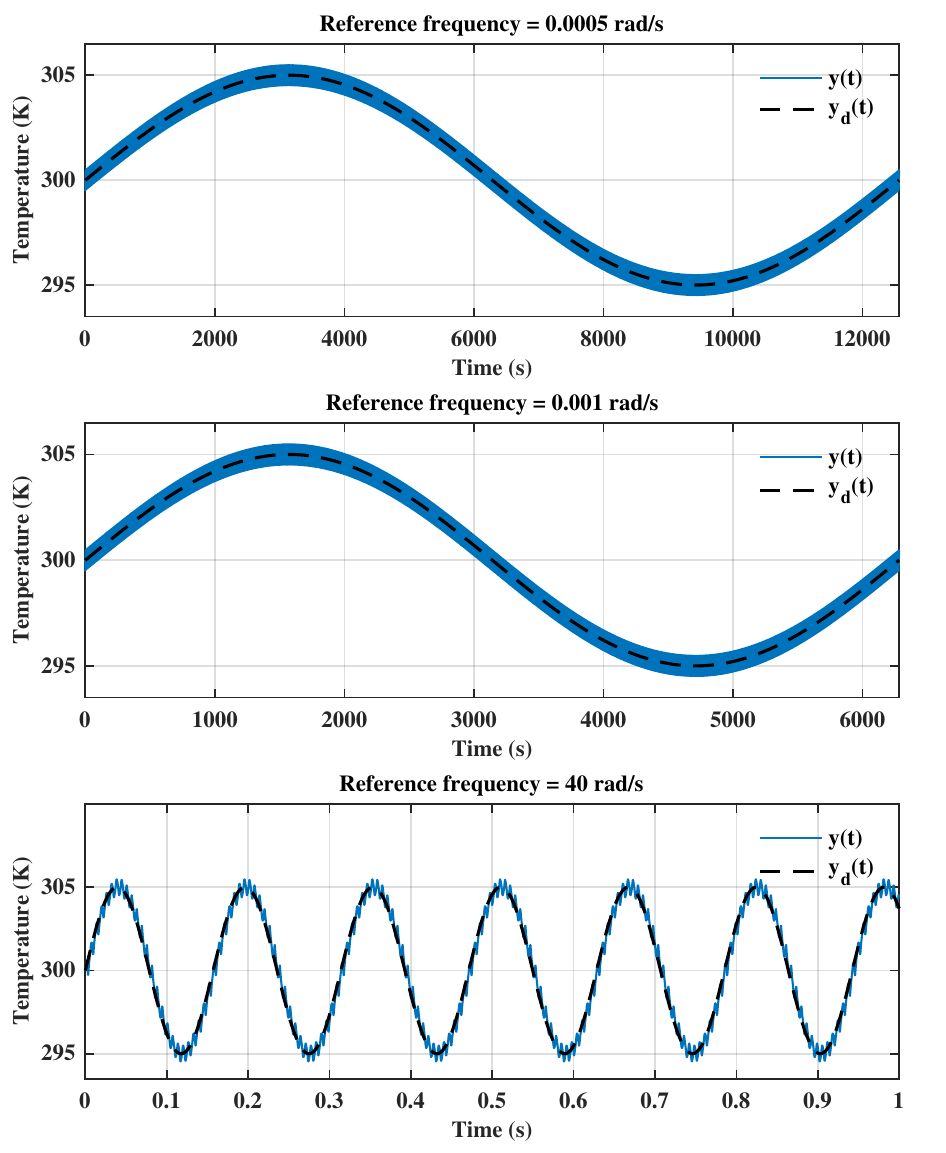}
        \caption{Controller~\eqref{eq:control_law}.}
        \label{fig:proposed_ESC_comparison}
    \end{subfigure}
    \caption{Temperature tracking at different reference frequencies using classical ESC and the proposed controller.}
    \label{fig:classical_ESC_comparison}
\end{figure}

Finally, we compare controller~\eqref{eq:control_law} with a classical ESC baseline~\cite{tan2006non}, obtained by setting $u=v$ in Figure \ref{fig:block_diagram}. The purpose is to determine how each controller responds to an increase in reference frequency once its tuning has been fixed. We use the reference $y_d(t)=300+5\sin(\nu t)$ in all tests, varying only its frequency $\nu$. Both controllers use the same initial conditions.

We first test classical ESC with a slowly varying reference at $\nu=0.0005\,\mathrm{rad/s}$. Classical ESC tracks this reference well with the selected tuning $(a,K,\omega)=(0.005,0.0005,0.01)$. We then increase $\nu$ to $0.001\,\mathrm{rad/s}$ without retuning. At this frequency, classical ESC loses tracking, and the simulation is stopped when the reactor temperature reaches $600\,\mathrm{K}$, as shown in Fig.~\ref{fig:classical_ESC_baseline}.

We repeat the tests at $\nu=0.0005$ and $0.001\,\mathrm{rad/s}$ with the proposed controller, using the fixed tuning $(a,\kappa,\omega)=(0.5,600,1000)$. The proposed controller maintains small tracking errors at both frequencies. With the same tuning, it also tracks the reference at $\nu=40\,\mathrm{rad/s}$, as shown in Fig.~\ref{fig:proposed_ESC_comparison}.

\subsection{Illustration of Vibrational Stabilisation}
\label{subsec:vibrational_stabilisation}

To illustrate the possibility of relaxing
Assumption~\ref{assum:Internal_dynamics}
through vibrational stabilisation, as discussed in
Remark~\ref{rem:Vibrational_control}, consider
\begin{equation}
    \begin{aligned}
        \dot{\eta} &= \bigl[\beta-c\sin^2(\nu y)\bigr]\eta,\\
        \dot{y} &= -u,
    \end{aligned}
    \label{eq:vibrational_example}
\end{equation}
where $\beta=9.6$, $c=20$, and $\nu=10\pi$.
The system has relative degree one, but its zero
dynamics, $\dot{\eta}=9.6\eta$, are unstable.
Consequently, the internal dynamics do not satisfy
Assumption~\ref{assum:Internal_dynamics}.
Nevertheless, high-frequency dither may stabilise
the internal dynamics, allowing output tracking
despite the instability of the unexcited zero
dynamics.
We select the time-varying reference trajectory
$y_d(t)=1+\sin(5t)$ and apply the proposed
controller~\eqref{eq:control_law} with
$\lambda=1$, $\kappa=220$, and
$\omega=1000\,\mathrm{rad/s}$.
Two dither amplitudes, $a=0.015$ and $a=0.04$,
are considered with identical initial conditions
$\eta(0)=1$, $y(0)=0.975$, and $\hat v(0)=0$.

\begin{figure}[!htb]
    \centering
    \begin{subfigure}[b]{0.49\linewidth}
        \centering
        \includegraphics[width=\linewidth]{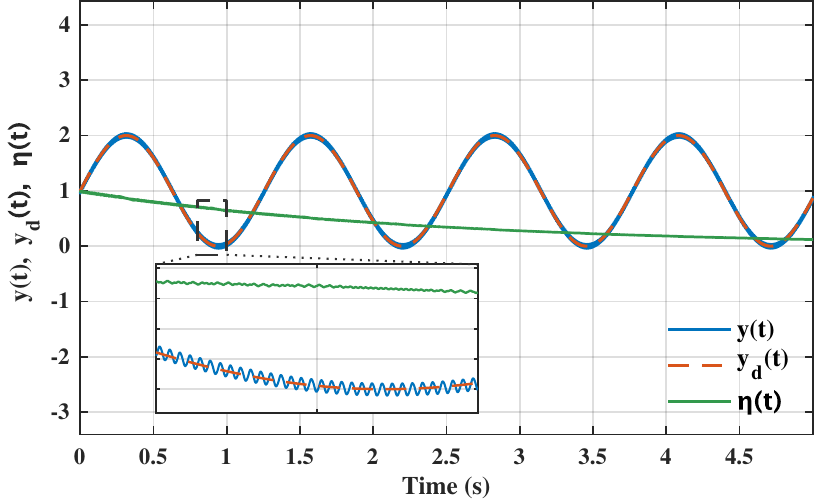}
        \caption{$a=0.04$}
        \label{fig:vibrational_stable}
    \end{subfigure}\hfill
    \begin{subfigure}[b]{0.49\linewidth}
        \centering
        \includegraphics[width=\linewidth]{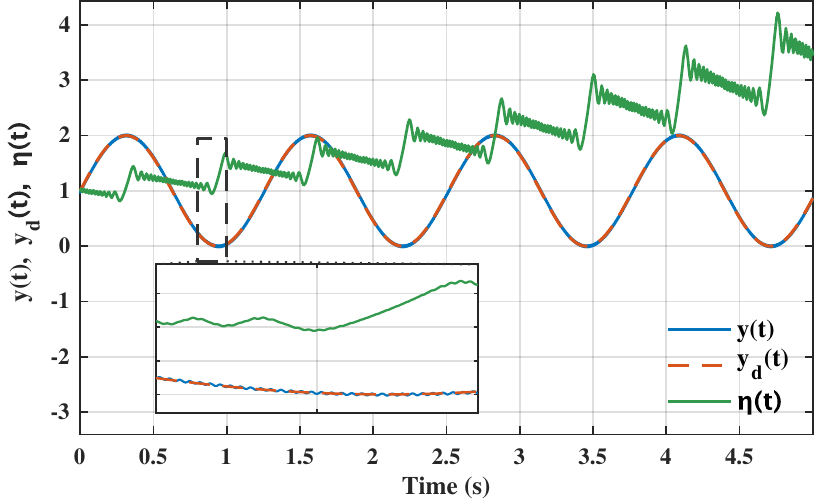}
        \caption{$a=0.015$}
        \label{fig:vibrational_unstable}
    \end{subfigure}
    \caption{Effect of dither amplitude on output tracking and vibrational stabilisation.}
\end{figure}

Figure~\ref{fig:vibrational_stable} shows that,
for $a=0.04$, the output closely tracks the
reference trajectory, while the internal state decays with small high-frequency oscillations.
In contrast, for $a=0.015$, the output maintains accurate tracking, but the internal state diverges, as shown in
Fig.~\ref{fig:vibrational_unstable}. 
These results illustrate that
Assumption~\ref{assum:Internal_dynamics}
may be relaxed when the internal dynamics
are vibrationally stabilisable.
However, vibrational stabilisation may require
a sufficiently large dither amplitude, whereas
the practical tracking-error bound in
Theorem~\ref{thm:tracking_SPES_transfer}
relies on selecting a sufficiently small amplitude.
Consequently, such a relaxation may weaken the
tracking guarantee, as the ultimate tracking
bound can no longer necessarily be prescribed
arbitrarily small.

\subsection{General Time-varying Objective Function}
\label{subsec:time_varying_objective}
We illustrate Remark~\ref{rem:time-varying cost} by considering the optimisation of a general time-varying objective function without plant dynamics. Specifically, consider
\begin{equation}
    J(t,u)
    =\frac12[1+0.5\sin(0.7t)]e^2+\frac14e^4,
    \label{eq:static_objective_example}
\end{equation}
where $e=u-y_d(t)$ and $y_d(t)=1+0.5\sin(0.5t)$. The objective function has a time-varying minimiser and
time-varying curvature. Moreover,
\[
    \frac{\partial J}{\partial u}
    =[1+0.5\sin(0.7t)]e+e^3,
\]
and
\[
    \frac{\partial^2 J}{\partial u^2}
    =1+0.5\sin(0.7t)+3e^2\geq 0.5.
\]
Thus, $J(t,u)$ is uniformly strongly convex,
with the unique global minimiser $u=y_d(t)$.

Since there are no plant dynamics, the controller~\eqref{eq:control_law} is modified to
\begin{equation}
    \begin{aligned}
        \dot{\hat v}&=-KJ(t,u)\sin(\omega t),\\
        v&=\hat v+a\sin(\omega t),\\
        u&=\lambda v.
    \end{aligned}
    \label{eq:static_objective_controller}
\end{equation}
For $\lambda=1$, averaging yields a gradient-like update law with an effective gain $\kappa=Ka/2$. We select
\[
    \lambda=1,\qquad a=0.01,\qquad
    \omega=400,\qquad K=2000,
\]
giving $\kappa=10$, with the initial condition $\hat v(0)=u(0)=0.9$.

\begin{figure}[h]
    \centering
    \includegraphics[width=0.8\linewidth]{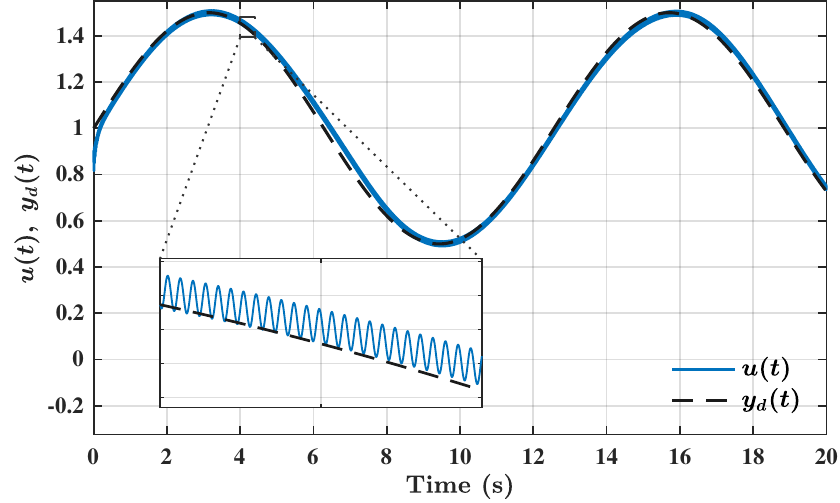}
    \caption{Tracking of a time-varying minimiser without plant dynamics. The inset shows the input and minimiser over $t\in[4,4.4]$.}
    \label{fig:static_objective_tracking}
\end{figure}

Figure~\ref{fig:static_objective_tracking} shows that $u(t)$ closely tracks the time-varying minimiser after an initial transient, with small dither-induced oscillations. This example illustrates that the proposed high-frequency-dither mechanism can be extended to general time-varying optimisation using only objective-value measurements, without requiring gradient information or an underlying dynamical plant.

\section{Conclusion}

This paper developed a high-frequency-dither-based
approach for model-free output tracking of unknown
nonlinear systems with relative degree one.
By exploiting the interaction between high-frequency
excitation and nonlinear plant dynamics, the proposed
controller achieves tracking without knowledge of
the plant model or control direction.
An averaging-based analysis established semiglobal
practical exponential tracking of time-varying
references. Future work will investigate extensions
to higher-relative-degree systems, more general
performance objectives, and input and output
constraints.

% \begin{ack}                               % Place acknowledgements
% Partially supported by the Roman Senate.  % here.
% \end{ack}

\bibliographystyle{plain}        % Include this if you use bibtex 
\bibliography{autosam}           % and a bib file to produce the 

\appendix
\section{Proof of Theorem \ref{thm:tracking_SPES_transfer}}
\label{Proof of Theorem}
The proof of Theorem~\ref{thm:tracking_SPES_transfer} is organised into two lemmas. First, we derive the partially averaged closed-loop dynamics in the Byrnes--Isidori coordinates defined by $\bm{\Theta}$ in \eqref{eq:BI_transformation} and establish boundedness and semiglobal practical exponential tracking under Assumption~\ref{assum:Internal_dynamics}.Then, we transfer the tracking property to the transformed closed-loop system, from which the physical output-tracking guarantee follows.

\subsection{Coordinate Transformation $\bm{\varphi}(s,\bm{\xi})$ using \eqref{eq:flow_transformation}}
%
%%%%%%%%%%%%%%%%%%%%%%%5
\iffalse
Consider the flow $\bm{\varphi}$ generated by $\bm{g}$, satisfying
\begin{equation}
    \label{eq:g_flow_derivative}
    \frac{\partial\bm{\varphi}}{\partial s}(s,\bm{\xi})
    =\bm{g}\bigl(\bm{\varphi}(s,\bm{\xi})\bigr),
    \qquad
    \bm{\varphi}(0,\bm{\xi})=\bm{\xi}.
\end{equation}
Let
\begin{equation}
    s(t)\coloneqq a\sin(\omega t),
    \label{eq:definition_s}
\end{equation}
and define the coordinate transformation
\begin{equation}
    \bm{x}=\bm{\varphi}(s,\bm{\xi}).
    \label{eq:g_flow_transformation}
\end{equation}
Since $\bm{g}$ is of class $C^2$, for any compact set $\mathcal N\subset\mathbb R^n$, there exists $\overline a>0$ such that the flow is defined on a neighbourhood of $[-\overline a,\overline a]\times\mathcal N$. For each fixed $|s|\leq\overline a$, the map $\bm{\varphi}(s,\cdot)$ is a diffeomorphism onto its image.

The flow identity
\begin{equation}
    \bm{\varphi}\bigl(s,\bm{\varphi}(r,\bm{\xi})\bigr)
    =\bm{\varphi}(s+r,\bm{\xi})
    \label{eq:flow_property}
\end{equation}
holds wherever both sides are defined. Thus, the inverse of \eqref{eq:g_flow_transformation} is
\[
    \bm{\xi}=\bm{\varphi}(-s,\bm{x}).
\]
In particular, $\bm{\xi}\to\bm{x}$ as $a\to0$, uniformly on compact sets.
%
\fi

Differentiating \eqref{eq:flow_transformation}
with respect to time and using
\eqref{eq:flow_definition} gives
\begin{equation}
    \dot{\bm{x}}
    =
    \nabla_{\bm{\xi}}\bm{\varphi}(s,\bm{\xi})
    \dot{\bm{\xi}}
    +
    \bm{g}\bigl(\bm{\varphi}(s,\bm{\xi})\bigr)
    a\omega\cos(\omega t).
    \label{eq:x_dot_transformation}
\end{equation}

Substituting \eqref{eq:flow_transformation}
into \eqref{eq:closed_loop_system} and comparing
with \eqref{eq:x_dot_transformation}, the terms
involving $a\omega\cos(\omega t)$ cancel, yielding
\bea
            \dot{\bm{\xi}}
        =
        \left[\nabla_{\bm{\xi}}
        \bm{\varphi}(s,\bm{\xi})\right]^{-1}
        \left[
        \bm{f}\left(\bm{\varphi}(s,\bm{\xi})\right)+
        \bm{g}\left(\bm{\varphi}(s,\bm{\xi})\right)
        \gamma(\dot{\hat v},\hat v, s)
        \right], \hspace{0.3in}
        \label{eq:xi_intermediate}
\eea
where 
\beas 
\gamma(\dot{\hat v},\hat v,s)=\dot{\hat v}+\hat v+s.
\eeas
According to \eqref{eq:flow_definition}, the flow generated by $\bm{g}$ satisfies
the composition property
\begin{equation}
    \bm{\varphi}\bigl(s,\bm{\varphi}(r,\bm{\xi})\bigr)
    =
    \bm{\varphi}(s+r,\bm{\xi}),
    \label{eq:flow_property}
\end{equation}
wherever both sides are defined.
Differentiating with respect to $r$ at $r=0$
and using \eqref{eq:flow_definition} yields
\begin{equation}
    \nabla_{\bm{\xi}}\bm{\varphi}(s,\bm{\xi})
    \bm{g}(\bm{\xi})
    =
    \bm{g}\bigl(\bm{\varphi}(s,\bm{\xi})\bigr).
    \label{eq:g_flow_pushforward}
\end{equation}

Since $\nabla_{\bm{\xi}}\bm{\varphi}(s,\bm{\xi})$
is nonsingular, substituting
\eqref{eq:g_flow_pushforward} into
\eqref{eq:xi_intermediate} yields
\eqref{eq:transformed_system}. In the transformed coordinates, define the tracking error as
\bea
    e_\xi \coloneqq h(\bm{\xi}) - y_d.
    \label{eq:e_xi}
\eea
From \eqref{eq:flow_definition}--\eqref{eq:flow_transformation},
the physical state satisfies
\[
\bm{x}=\bm \varphi(a\sin(\omega t),\bm{\xi}),
\]
where $\bm \varphi(0,\bm{\xi})=\bm{\xi}$.
Since $|a\sin(\omega t)|\leq a$, the flow transformation
approaches the identity mapping as $a\to 0$.
Consequently, by continuity of $h$,
\[
e-e_\xi
=h\bigl(\bm \varphi(a\sin(\omega t),\bm{\xi})\bigr)
-h(\bm{\xi})\to 0.
\]
Thus, for sufficiently small dither amplitudes,
the transformed tracking error $e_\xi$ approximates
the physical tracking error $e$.

\subsection{Byrnes--Isidori Coordinate Transformation}
\label{sec:BI_before_averaging}

We now express the transformed closed-loop system
\eqref{eq:transformed_system} in Byrnes--Isidori
coordinates before applying partial averaging.
For convenience, define
\begin{equation}
    \begin{aligned}
        \mathcal J(t,s,\bm\xi)
        &\coloneqq
        J\bigl(t,h(\bm\varphi(s,\bm\xi))\bigr)\\
        &=
        \bigl[
            h(\bm\varphi(s,\bm\xi))-y_d(t)
        \bigr]^2.
    \end{aligned}
    \label{eq:transformed_objective_definition}
\end{equation}
With this notation and (\ref{eq:dither_signal}),
\eqref{eq:transformed_system} becomes
\begin{equation}
    \begin{aligned}
        \dot{\bm\xi}
        ={}&
        \bm f_1(s,\bm\xi)
        +\bm g(\bm\xi)
        \bigl(\dot{\hat v}+\hat v+s\bigr),\\
        \dot{\hat v}
        ={}&
        -K\mathcal J(t,s,\bm\xi)\sin(\omega t).
    \end{aligned}
    \label{eq:transformed_system_compact}
\end{equation}

Apply the global coordinate transformation
$\bm\Theta$ defined in \eqref{eq:BI_transformation},
specified in Assumption~\ref{assum:Internal_dynamics},
to the $\bm \varphi$-transformed state $\bm\xi$:
\begin{equation}
    \begin{bmatrix}
        \bm\eta\\
        y_\xi
    \end{bmatrix}
    =
    \bm\Theta(\bm\xi)
    =
    \begin{bmatrix}
        \bm\vartheta(\bm\xi)\\
        h(\bm\xi)
    \end{bmatrix}.
    \label{eq:BI_transformed_coordinates}
\end{equation}
Since the second component of $\bm\Theta(\bm\xi)$
is $h(\bm\xi)$, we have $y_\xi=h(\bm\xi)$.
Therefore, the transformed tracking error remains
as defined in \eqref{eq:e_xi}.

Define the phase-dependent functions
\begin{equation}
    \begin{aligned}
        \bm q_s(\bm\eta,y_\xi)
        \coloneqq{}&
        \left.
        \frac{\partial\bm\vartheta}{\partial\bm\xi}
        (\bm\xi)\bm f_1(s,\bm\xi)
        \right|_{\bm\xi=
        \bm\Theta^{-1}(\bm\eta,y_\xi)},\\
        \ell_s(\bm\eta,y_\xi)
        \coloneqq{}&
        \left.
        \frac{\partial h}{\partial\bm\xi}
        (\bm\xi)\bm f_1(s,\bm\xi)
        \right|_{\bm\xi=
        \bm\Theta^{-1}(\bm\eta,y_\xi)}.
    \end{aligned}
    \label{eq:BI_phase_dependent_drifts}
\end{equation}
Since $\bm\varphi(0,\bm\xi)=\bm\xi$ and
$\nabla_{\bm\xi}\bm\varphi(0,\bm\xi)=I$,
\eqref{eq:f1_definition} gives
\begin{equation}
    \bm f_1(0,\bm\xi)=\bm f(\bm\xi).
    \label{eq:f_1 property}
\end{equation}
Therefore,
\begin{equation}
    \begin{aligned}
        \bm q_0(\bm\eta,y_\xi)
        &=\bm q(\bm\eta,y_\xi),\\
        \ell_0(\bm\eta,y_\xi)
        &=\ell(\bm\eta,y_\xi),
    \end{aligned}
    \label{eq:BI_zero_phase_drifts}
\end{equation}
where $\bm q$ and $\ell$ come from \eqref{eq:q_definition} and \eqref{eq:ell_definition}, respectively.

Differentiating $\bm\eta=\bm\vartheta(\bm\xi)$
using \eqref{eq:transformed_system_compact}
and \eqref{eq:BI_eta_condition_x} yields
\bea
    \dot{\bm\eta}
    &=&
    \frac{\partial\bm\vartheta}{\partial\bm\xi}
    (\bm\xi)\dot{\bm\xi}
    \nonumber\\
    &=&
    \frac{\partial\bm\vartheta}{\partial\bm\xi}
    (\bm\xi)\bm f_1(s,\bm\xi)
    + \frac{\partial\bm\vartheta}{\partial\bm\xi}
    (\bm\xi)\bm g(\bm\xi)
    \bigl(\dot{\hat v}+\hat v+s\bigr)
    \nonumber\\
    &=&
    \bm q_s(\bm\eta,y_\xi).
    \label{eq:BI_unaveraged_internal_derivation}
\eea
Finally, differentiating \eqref{eq:e_xi} and substituting the controller dynamics \eqref{eq:transformed_system_compact},
the transformed closed-loop system in Byrnes--Isidori coordinates is
\begin{subequations}
\label{eq:unaveraged_BI_system}
\begin{align}
    \dot{\bm\eta}
    &=
    \bm q_s(\bm\eta,y_\xi),
    \label{eq:unaveraged_BI_eta}\\
    \dot e_\xi
    &=
    \ell_s(\bm\eta,y_\xi)
    +b(\bm\eta,y_\xi)(\hat v+s)
    \nonumber\\
    &\quad
    {}-Kb(\bm\eta,y_\xi)
    \mathcal J(t,s,\bm\xi)\sin(\omega t)
    -\dot y_d(t),
    \label{eq:unaveraged_BI_error}\\
    \dot{\hat v}
    &=
    -K\mathcal J(t,s,\bm\xi)\sin(\omega t).
    \label{eq:unaveraged_BI_v}
\end{align}
\end{subequations}
where $y_\xi=e_\xi+y_d(t)$ and $s$ comes from (\ref{eq:dither_signal}).

\subsection{Partially Averaged System}
\label{sec:partial_averaging_BI}

We now apply partial averaging to \eqref{eq:unaveraged_BI_system}. The averaging is performed over the fast timescale $\theta=\omega t$, holding $t$, $\bm\eta$, $e_\xi$, $y_d$ and $\hat v$ fixed. Consequently, $y_\xi$ and $\bm\xi$ are also fixed during the phase integration. For convenience, define
\bea
    \langle P\rangle_\theta
    &\coloneqq&
    \frac{1}{2\pi}
    \int_0^{2\pi}P(\theta)\,d\theta.
    \label{eq:phase_average_definition}
\eea

We first calculate the averages of $\bm q_s$ and $\ell_s$. Expanding $\bm f_1(s,\bm\xi)$ about $s=0$ using Taylor's theorem with integral remainder gives
\bea
    \bm f_1(s,\bm\xi)
    &=&
    \bm f(\bm\xi)
    +s\frac{\partial\bm f_1}{\partial s}
    (0,\bm\xi)
    \nonumber\\
    &&
    {}+s^2\int_0^1(1-\tau)
    \frac{\partial^2\bm f_1}{\partial s^2}
    (\tau s,\bm\xi)\,d\tau,
    \label{eq:F_Taylor_expansion}
\eea
where we have used $\bm f_1(0,\bm\xi)=\bm f(\bm\xi)$. Define
\bea
    \bm\Gamma_f(\bm\xi,a)
    &\coloneqq&
    \frac{1}{2\pi}
    \int_0^{2\pi}\sin^2\theta
    \int_0^1(1-\tau)
    \nonumber\\
    &&
    {}\times
    \frac{\partial^2\bm f_1}{\partial s^2}
    (\tau a\sin\theta,\bm\xi)
    \,d\tau\,d\theta.
    \label{eq:Gamma_f_definition}
\eea
Since the coordinate transformation $\bm\Theta$ is independent of the fast timescales $\omega t$ and $\bm\xi$ is fixed during the integration, the derivatives $\partial\bm\vartheta/\partial\bm\xi$ and $\partial h/\partial\bm\xi$ are also fixed. Multiplying \eqref{eq:F_Taylor_expansion} by these derivatives and using $\langle\sin\theta\rangle_\theta=0$ yields
\bea
    \bigl\langle
        \bm q_{a\sin\theta}(\bm\eta,y_\xi)
    \bigr\rangle_\theta
    &=&
    \bm q(\bm\eta,y_\xi)+\bm r_\eta,
    \nonumber\\
    \bigl\langle
        \ell_{a\sin\theta}(\bm\eta,y_\xi)
    \bigr\rangle_\theta
    &=&
    \ell(\bm\eta,y_\xi)+r_y,
    \label{eq:BI_averaged_drift_terms}
\eea
where the remainder is defined by
\bea
    \bm r_\eta(\bm\eta,y_\xi,a)
    &\coloneqq&
    a^2
    \frac{\partial\bm\vartheta}{\partial\bm\xi}
    (\bm\xi)\bm\Gamma_f(\bm\xi,a),
    \nonumber\\
    r_y(\bm\eta,y_\xi,a)
    &\coloneqq&
    a^2
    \frac{\partial h}{\partial\bm\xi}
    (\bm\xi)\bm\Gamma_f(\bm\xi,a),
    \label{eq:BI_drift_remainder_definitions}
\eea
with $\bm\xi=\bm\Theta^{-1}(\bm\eta,y_\xi)$ and $y_\xi = e_\xi + y_d(t)$.

Next, we calculate the averaged controller dynamics. Taylor expansion of $\mathcal J(t,s,\bm\xi)$ about $s=0$ gives
\bea
    \mathcal J(t,s,\bm\xi)
    &=&
    \mathcal J(t,0,\bm\xi)
    +s\frac{\partial\mathcal J}{\partial s}
    (t,0,\bm\xi)+\frac{s^2}{2}
    \frac{\partial^2\mathcal J}{\partial s^2}
    (t,0,\bm\xi)
    \nonumber\\
    &&
    {}+\frac{s^3}{2}
    \int_0^1(1-\tau)^2
    \frac{\partial^3\mathcal J}{\partial s^3}
    (t,\tau s,\bm\xi)\,d\tau.
    \label{eq:Q_Taylor_expansion}
\eea
By the definition of $\mathcal J$ and the flow equation \eqref{eq:flow_definition},
\bea
    \mathcal J(t,0,\bm\xi)
    &=&
    \bigl[h(\bm\xi)-y_d(t)\bigr]^2
    =e_\xi^2,
    \nonumber\\
    \frac{\partial\mathcal J}{\partial s}
    (t,0,\bm\xi)
    &=&
    2\bigl[h(\bm\xi)-y_d(t)\bigr]
    L_{\bm g}h(\bm\xi)
    \nonumber\\
    &=&
    2e_\xi b(\bm\eta,y_\xi).
    \label{eq:Q_derivative_at_zero}
\eea
The higher derivatives in \eqref{eq:Q_Taylor_expansion} exist under the assumed regularity of $\bm g$ and $h$.

Substituting $s=a\sin\theta$ into \eqref{eq:Q_Taylor_expansion}, multiplying by $\sin\theta$ gives
\bea
    &&
    \bigl\langle
        \mathcal J(t,a\sin\theta,\bm\xi)
        \sin\theta
    \bigr\rangle_\theta
    \nonumber\\
    &=&
    \frac a2
    \frac{\partial\mathcal J}{\partial s}
    (t,0,\bm\xi)
    +\frac{a^3}{2}\Gamma_v(t,\bm\xi,a),
    \label{eq:Q_phase_average}
\eea
where
\bea
    \Gamma_v(t,\bm\xi,a)
    &\coloneqq&
    \frac{1}{2\pi}
    \int_0^{2\pi}\sin^4\theta
    \int_0^1(1-\tau)^2
    \nonumber\\
    &&
    {}\times
    \frac{\partial^3\mathcal J}{\partial s^3}
    (t,\tau a\sin\theta,\bm\xi)
    \,d\tau\,d\theta.
    \label{eq:Gamma_v_definition}
\eea
As $\kappa=Ka/2$, using  \eqref{eq:Q_derivative_at_zero} and \eqref{eq:Q_phase_average} yields
\bea
    &&
    \bigl\langle
        -K\mathcal J(t,a\sin\theta,\bm\xi)
        \sin\theta
    \bigr\rangle_\theta
    \nonumber\\
    &=&
    -\kappa
    \frac{\partial\mathcal J}{\partial s}
    (t,0,\bm\xi)
    -\kappa a^2\Gamma_v(t,\bm\xi,a)
    \nonumber\\
    &=&
    -2\kappa b(\bm\eta,y_\xi)e_\xi+r_v,
    \label{eq:BI_controller_phase_average}
\eea
where
\bea
    r_v(t,\bm\eta,y_\xi,a,\kappa)
    &\coloneqq&
    -\kappa a^2\Gamma_v(t,\bm\xi,a).
    \label{eq:BI_controller_remainder}
\eea
Moreover, $b(\bm\eta,y_\xi)$ is independent of the fast timescale $\theta$ at the fixed state, so
\bea
    \bigl\langle
        b(\bm\eta,y_\xi)a\sin\theta
    \bigr\rangle_\theta
    &=&
    0.
    \label{eq:BI_additive_dither_average}
\eea
Let $F_e(t,\bm\eta,e_\xi,\hat v,\theta)$ denote the right-hand side of the tracking-error equation in \eqref{eq:unaveraged_BI_system}, with $s=a\sin\theta$ and $\sin(\omega t)$ replaced by $\sin\theta$.
Direct partial averaging of this error dynamics, and by \eqref{eq:BI_averaged_drift_terms}, \eqref{eq:BI_controller_phase_average}, and \eqref{eq:BI_additive_dither_average}, we have
\bea
    &{}&\bigl\langle F_e(t,\bm\eta,e_\xi,\hat v,\theta)\bigr\rangle_\theta
    \nonumber \\
    &\coloneqq&
    \bigl\langle
        \ell_{a\sin\theta}(\bm\eta,y_\xi)
    \bigr\rangle_\theta
    +b(\bm\eta,y_\xi)\hat v 
    +
    \bigl\langle
        b(\bm\eta,y_\xi)a\sin\theta
    \bigr\rangle_\theta
    \nonumber\\
    &&
    -\dot y_d(t)
    +
    b(\bm\eta,y_\xi)
    \bigl\langle
        -K\mathcal J(t,a\sin\theta,\bm\xi)
        \sin\theta
    \bigr\rangle_\theta
    \nonumber\\
    &=&
    -2\kappa b^2(\bm\eta,y_\xi)e_\xi
    +\ell(\bm\eta,y_\xi)
    \nonumber\\
    &&
    {}+b(\bm\eta,y_\xi)\hat v
    -\dot y_d(t)
    +b(\bm\eta,y_\xi)r_v+r_y.
    \label{eq:direct_average_error_vector_field}
\eea
Thus the partial averaged system of \eqref{eq:unaveraged_BI_system} is given by
\begin{subequations}
\label{eq:averaged_BI_system}
\begin{align}
    \dot{\bm\eta}^{av}
    &=
    \bm q(\bm\eta^{av},y_\xi^{av})
    +\bm r_\eta,
    \label{eq:averaged_BI_eta}\\
    \dot e_\xi^{av}
    &=
    -2\kappa b^2(\bm\eta^{av},y_\xi^{av})e_\xi^{av} +\ell(\bm\eta^{av},y_\xi^{av})
    \nonumber\\
    &\quad
    {}
    +b(\bm\eta^{av},y_\xi^{av})\hat v^{av}
    -\dot y_d(t)
    \nonumber\\
    &\quad
    {}+b(\bm\eta^{av},y_\xi^{av})r_v+r_y,
    \label{eq:averaged_BI_error}\\
    \dot{\hat v}^{av}
    &=
    -2\kappa b(\bm\eta^{av},y_\xi^{av})e_\xi^{av}
    +r_v,
    \label{eq:averaged_BI_v}
\end{align}
\end{subequations}
where $b(\bm \eta^{av}, y_\xi^{av}) \neq 0$ by Assumption~\ref{assum:Relative_degree}.

The remainder functions in the partial averaged system \eqref{eq:averaged_BI_system} (i.e., \eqref{eq:BI_drift_remainder_definitions} and \eqref{eq:BI_controller_remainder}) admit uniform bounds on compact state sets. Indeed, let $\mathcal P$ be a compact set in $(\bm\eta,y_\xi)$ coordinates and define $\mathcal N=\bm\Theta^{-1}(\mathcal P)$. Choose $\bar a>0$ sufficiently small that the flow is defined on a neighbourhood of $[-\bar a,\bar a]\times\mathcal N$. The derivatives appearing in \eqref{eq:Gamma_f_definition} and \eqref{eq:Gamma_v_definition} are bounded there, uniformly in $t$ by Assumption~\ref{assum:Reference_trajectory}. The coordinate derivatives are also bounded on $\mathcal N$. Consequently, there exists $C>0$, independent of $t$, $a$, and $\kappa$, such that
\bea
    |\bm r_\eta(\bm\eta,y_\xi,a)|
    +|r_y(\bm\eta,y_\xi,a)|
    &\leq&
    Ca^2,
    \nonumber\\
    |r_v(t,\bm\eta,y_\xi,a,\kappa)|
    &\leq&
    C\kappa a^2
    \label{eq:BI_remainder_bounds}
\eea
for all $(\bm\eta,y_\xi)\in\mathcal P$, $t\geq0$, $0<a\leq\bar a$, and $\kappa>0$. 

\subsection{Stability guarantees}

We first establish semiglobal practical exponential
stability (SPES) of the averaged tracking error
$e_\xi^{av}$ for the partially averaged system
\eqref{eq:averaged_BI_system}. We then extend
this result to the transformed tracking error
$e_\xi$ and, finally, to the original tracking
error $e$.

\begin{lem}[SPES of the averaged tracking error]
\label{lem:averaged_tracking_SPAS}
Suppose Assumptions~\ref{assum:Relative_degree}--\ref{assum:Reference_trajectory} hold.
For any prescribed $\Delta^{av},\delta^{av},\rho>0$,
there exist constants $\kappa^*>0$ and $a^*>0$
such that, for every $\kappa\geq\kappa^*$ and
$0<a\leq a^*$, all solutions of the partially
averaged system \eqref{eq:averaged_BI_system}
satisfying
\[
\left|
\begin{bmatrix}
\bm\xi^{av}(t_0)\\
\hat v^{av}(t_0)
\end{bmatrix}
\right|\leq\Delta^{av}
\]
are forward complete, with all averaged states
remaining bounded, and the tracking error satisfying
\begin{equation}
|e_\xi^{av}(t)|
\leq e^{-\rho(t-t_0)}|e_\xi^{av}(t_0)|
+\delta^{av},
\label{eq:averaged_tracking_SPAS}
\end{equation}
for all $t\geq t_0$.
\hfill$\square$
\end{lem}

\noindent\textbf{Proof:} The proof of Lemma~\ref{lem:averaged_tracking_SPAS} is given in Appendix~\ref{Proof of Lemma}.

We now use Lemmas~\ref{lem:averaged_tracking_SPAS} to establish
the SPES property of the transformed tracking
error $e_\xi$ defined in \eqref{eq:e_xi}.

\begin{lem}[SPES of $e_\xi$]
    \label{lem:SPES_of_e_xi}
    Suppose Assumptions~\ref{assum:Relative_degree}--\ref{assum:Reference_trajectory} hold. For any $\Delta_\xi,\delta_\xi,\rho_\xi>0$, there exist $\kappa_\xi^*>0$ and $a_\xi^*>0$ such that, for all $\kappa\geq\kappa_\xi^*$ and $0<a\leq a_\xi^*$, there exists $\omega^*>0$ such that, for every $\omega\geq\omega^*$, every solution of \eqref{eq:transformed_system} with $|(\bm{\xi}(t_0),\hat v(t_0))|\leq\Delta_\xi$ is forward complete, remains bounded, and satisfies
    \begin{equation}
        |e_\xi(t)|
        \leq e^{-\rho_\xi(t-t_0)}|e_\xi(t_0)|+\delta_\xi,
                \label{eq:shifted_system_SPAS}
    \end{equation}
for all $t\geq t_0$.
    \hfill$\square$
\end{lem}

\noindent\textbf{Proof:} The proof of Lemma~\ref{lem:SPES_of_e_xi} is given in Appendix~\ref{Proof of Lemma e_xi}.
% By Lemma~\ref{lem:averaged_tracking_SPAS},
% the averaged tracking error $e_\xi^{av}$
% satisfies the SPES estimate, and the averaged
% trajectories remain in a compact set for
% sufficiently large $\kappa$ and sufficiently
% small $a$.

% For each fixed admissible pair $(\kappa,a)$,
% the regularity conditions of
% Lemma~\ref{lem:partial_averaging}
% hold on the corresponding compact neighbourhood.
% Applying this lemma establishes boundedness,
% forward completeness, and the tracking estimate
% \eqref{eq:shifted_system_SPAS}
% for sufficiently large $\omega$.
% \hfill$\blacksquare$

\medskip
\noindent\textbf{Proof of
Theorem~\ref{thm:tracking_SPES_transfer}:}
Fix $\Delta,\rho,\delta>0$.
By \eqref{eq:flow_transformation}, for sufficiently
small $a$, the initial condition
$|(\bm{x}(t_0),\hat v(t_0))|\leq\Delta$
corresponds to
$|(\bm{\xi}(t_0),\hat v(t_0))|\leq\Delta_\xi$
for some $\Delta_\xi>0$, uniformly in
$t_0$ and $\omega$.

Applying Lemma~\ref{lem:SPES_of_e_xi}
with $\rho_\xi=\rho$ and $\delta_\xi=\delta/2$
gives the SPES estimate for $e_\xi$.

On the compact set containing the transformed
trajectories, smoothness of $h$ and
$\bm{\varphi}$ implies
\[
\begin{aligned}
|e(t)-e_\xi(t)|
&=\bigl|h\bigl(
\bm{\varphi}(a\sin(\omega t),\bm{\xi}(t))
\bigr)-h(\bm{\xi}(t))\bigr|\\
&\leq C_h a,
\end{aligned}
\]
where $C_h>0$ is independent of $a$ and
$\omega$ on the chosen compact set.

Choosing $a$ sufficiently small such that
$C_h a\leq\delta/4$, we obtain
\[
\begin{aligned}
|e(t)|
&\leq |e_\xi(t)|+C_h a\\
&\leq e^{-\rho(t-t_0)}|e_\xi(t_0)|
+\frac{\delta}{2}+C_h a\\
&\leq e^{-\rho(t-t_0)}|e(t_0)|
+\frac{\delta}{2}+2C_h a\\
&\leq e^{-\rho(t-t_0)}|e(t_0)|+\delta,
\qquad t\geq t_0.
\end{aligned}
\]

Moreover, Lemma~\ref{lem:SPES_of_e_xi}
ensures boundedness of the transformed
trajectories $(\bm\xi(t),\hat v(t))$.
Since
\[
\bm{x}(t)=
\bm{\varphi}(a\sin(\omega t),\bm\xi(t)),
\]
and $|a\sin(\omega t)|\leq a$,
the smoothness of the flow transformation
implies boundedness of $\bm{x}(t)$.
Consequently, all closed-loop states
remain bounded, and the original
closed-loop system is forward complete.
This completes the proof.
\hfill$\blacksquare$

\section{Proof of Lemma \ref{lem:averaged_tracking_SPAS}}
\label{Proof of Lemma}

Define
\begin{equation}
    H(\bm{\eta},y_\xi)
    \coloneqq\int_0^{y_\xi}\frac{dr}{b(\bm{\eta},r)}.
    \label{eq:tracking_H}
\end{equation}
Since $b$ is continuously differentiable and nowhere zero,
$H$ is continuously differentiable, with
\[
    \begin{aligned}
        \frac{\partial H}{\partial y_\xi}(\bm{\eta},y_\xi)
        &=\frac{1}{b(\bm{\eta},y_\xi)},\\
        \frac{\partial H}{\partial\bm{\eta}}(\bm{\eta},y_\xi)
        &=-\int_0^{y_\xi}
        \frac{1}{b(\bm{\eta},r)^2}
        \frac{\partial b}{\partial\bm{\eta}}(\bm{\eta},r)\,dr.
    \end{aligned}
\]
Introduce
\[
    \begin{aligned}
        w^{av}
        &\coloneqq\hat v^{av}
        -H(\bm{\eta}^{av},y_\xi^{av}),\\
        \hat v^{av}
        &=w^{av}+H(\bm{\eta}^{av},y_\xi^{av}).
    \end{aligned}
\]
Differentiating $w^{av}$ along
\eqref{eq:averaged_BI_system} gives
\[
    \begin{aligned}
        \dot w^{av}
        &=\dot{\hat v}^{av}
        -\frac{\partial H}{\partial\bm{\eta}}
        \dot{\bm{\eta}}^{av}
        -\frac{1}{b}\dot y_\xi^{av}\\
        &=-\hat v^{av}-\frac{\ell}{b}
        -\frac{\partial H}{\partial\bm{\eta}}\bm{q}
        -\frac{\partial H}{\partial\bm{\eta}}\bm{r}_\eta
        -\frac{r_y}{b}\\
        &=-w^{av}
        -\mathcal B(\bm{\eta}^{av},y_\xi^{av})+r_w,
    \end{aligned}
\]
where the functions in the first two lines are evaluated
along the averaged trajectory, and
\[
    \begin{aligned}
        \mathcal B(\bm{\eta},y_\xi)
        &\coloneqq H(\bm{\eta},y_\xi)
        +\frac{\ell(\bm{\eta},y_\xi)}{b(\bm{\eta},y_\xi)}\\
        &\quad
        +\frac{\partial H}{\partial\bm{\eta}}
        (\bm{\eta},y_\xi)\bm{q}(\bm{\eta},y_\xi),\\
        r_w(\bm{\eta},y_\xi,a)
        &\coloneqq
        -\frac{\partial H}{\partial\bm{\eta}}
        (\bm{\eta},y_\xi)\bm{r}_\eta(\bm{\eta},y_\xi,a)\\
        &\quad
        -\frac{r_y(\bm{\eta},y_\xi,a)}
        {b(\bm{\eta},y_\xi)}.
    \end{aligned}
\]
In particular, the feedback term
$-2\kappa b e_\xi^{av}$ and the controller remainder
$r_v$ cancel from the equation for $\dot w^{av}$.
Let
\[
    \begin{aligned}
        \mathcal A(\bm{\eta},y_\xi,w)
        &\coloneqq\ell(\bm{\eta},y_\xi)\\
        &\quad
        +b(\bm{\eta},y_\xi)
        \bigl[w+H(\bm{\eta},y_\xi)\bigr].
    \end{aligned}
\]
The transformed averaged system is
\begin{equation}
    \begin{aligned}
        \dot{\bm{\eta}}^{av}
        ={}&\bm{q}(\bm{\eta}^{av},y_\xi^{av})
        +\bm{r}_\eta,\\
        \dot w^{av}
        ={}&-w^{av}
        -\mathcal B(\bm{\eta}^{av},y_\xi^{av})+r_w,\\
        \dot e_\xi^{av}
        ={}&-2\kappa b^2(\bm{\eta}^{av},y_\xi^{av})
        e_\xi^{av}\\
        &+\mathcal A(\bm{\eta}^{av},y_\xi^{av},w^{av})
        -\dot y_d(t)\\
        &+b(\bm{\eta}^{av},y_\xi^{av})r_v+r_y,
    \end{aligned}
    \label{eq:tracking_transformed_system}
\end{equation}
where the remainder functions are evaluated along the
averaged trajectory at the indicated parameters.
The error equation holds almost everywhere.

For the prescribed $\Delta^{av}>0$, define
\[
    \mathcal K_{\Delta^{av}}
    \coloneqq
    \bigl\{(\bm{\xi},\hat v):
    |(\bm{\xi},\hat v)|\leq\Delta^{av}\bigr\}.
\]
Let
\[
    \begin{aligned}
        E_0
        &\coloneqq R+
        \max_{(\bm{\xi},\hat v)\in\mathcal K_{\Delta^{av}}}
        |h(\bm{\xi})|,\\
        W_0
        &\coloneqq
        \max_{(\bm{\xi},\hat v)\in\mathcal K_{\Delta^{av}}}
        W(\bm{\vartheta}(\bm{\xi})),\\
        L_0
        &\coloneqq
        \max_{(\bm{\xi},\hat v)\in\mathcal K_{\Delta^{av}}}
        \bigl|\hat v
        -H(\bm{\vartheta}(\bm{\xi}),h(\bm{\xi}))\bigr|.
    \end{aligned}
\]
These constants bound $|e_\xi^{av}(t_0)|$,
$W(\bm{\eta}^{av}(t_0))$, and $|w^{av}(t_0)|$,
respectively.

Choose $E\geq E_0$ with $E>0$, and let
\begin{equation}
    Y\coloneqq R+E.
    \label{eq:tracking_Y}
\end{equation}
Since $\alpha\in\mathcal K_\infty$, choose
$c_W\geq W_0$ such that
\begin{equation}
    \alpha(c_W)\geq\sigma(Y)+c_0+2.
    \label{eq:tracking_cW_choice}
\end{equation}
Properness of $W$ implies that
\[
    \mathcal P
    \coloneqq
    \bigl\{(\bm{\eta},y_\xi):
    W(\bm{\eta})\leq c_W,\ |y_\xi|\leq Y\bigr\}
\]
is compact. Define
\[
    B_{\max}\coloneqq
    \max_{(\bm{\eta},y_\xi)\in\mathcal P}
    |\mathcal B(\bm{\eta},y_\xi)|,
\]
and choose $L\geq L_0$ such that
\begin{equation}
    L\geq B_{\max}+2.
    \label{eq:tracking_L_choice}
\end{equation}

Consider the compact set
\begin{equation}
    \begin{aligned}
        \mathcal S\coloneqq
        \bigl\{(\bm{\eta},w,e_\xi):\
        &W(\bm{\eta})\leq c_W,\\
        &|w|\leq L,\ |e_\xi|\leq E\bigr\}.
    \end{aligned}
    \label{eq:tracking_S}
\end{equation}
Every admissible initial condition, expressed in these
coordinates, lies in $\mathcal S$.
Whenever $(\bm{\eta}^{av},w^{av},e_\xi^{av})\in\mathcal S$,
\eqref{eq:tracking_Y} and $|y_d(t)|\leq R$ imply
$(\bm{\eta}^{av},y_\xi^{av})\in\mathcal P$.

Define
\begin{equation}
    \begin{aligned}
        b_{\min}
        &\coloneqq
        \min_{(\bm{\eta},y_\xi)\in\mathcal P}
        |b(\bm{\eta},y_\xi)|>0,\\
        M
        &\coloneqq
        \max_{\substack{(\bm{\eta},y_\xi)\in\mathcal P\\
                        |w|\leq L}}
        |\mathcal A(\bm{\eta},y_\xi,w)|,\\
        \nu&\coloneqq2b_{\min}^2.
    \end{aligned}
    \label{eq:tracking_compact_constants}
\end{equation}
The sets $\mathcal P,\mathcal S$ and these constants
are chosen independently of $\kappa$ and $a$.

Choose $\bar a>0$ sufficiently small that
$\bm{\varphi}(s,\bm{\xi})$ is defined on a neighbourhood
of $[-\bar a,\bar a]\times\bm{\Theta}^{-1}(\mathcal P)$.
The remainders defined in
\eqref{eq:BI_drift_remainder_definitions} and
\eqref{eq:BI_controller_remainder} satisfy
\eqref{eq:BI_remainder_bounds}.
Together with compactness of $\mathcal P$ and
continuity of $\partial W/\partial\bm{\eta}$,
$\partial H/\partial\bm{\eta}$, $b$, and $1/b$,
these bounds imply that there exists $C>0$,
independent of $\kappa$, $a$, and $t$, such that
\begin{equation}
    \begin{aligned}
        \left|
        \frac{\partial W}{\partial\bm{\eta}}\bm{r}_\eta
        \right|&\leq Ca^2,\\
        |r_w|&\leq Ca^2,\\
        |r_y|&\leq Ca^2,\\
        |b r_v|&\leq C\kappa a^2
    \end{aligned}
    \label{eq:tracking_remainder_bounds}
\end{equation}
for all $(\bm{\eta},y_\xi)\in\mathcal P$, $t\geq0$,
$0<a\leq\bar a$, and $\kappa>0$.

Let
\begin{equation}
    C_0\coloneqq M+D+C\bar a^2.
    \label{eq:tracking_C0}
\end{equation}
Whenever $(\bm{\eta}^{av},w^{av},e_\xi^{av})\in\mathcal S$,
\eqref{eq:tracking_transformed_system} and
\eqref{eq:tracking_compact_constants}--\eqref{eq:tracking_C0}
imply, for almost every $t$,
\begin{equation}
    D^+|e_\xi^{av}|
    \leq-\nu\kappa|e_\xi^{av}|+C_0+C\kappa a^2.
    \label{eq:tracking_error_inequality}
\end{equation}
Here $D^+$ denotes the upper right Dini derivative.

Define
\begin{equation}
    \begin{aligned}
        a_0
        &\coloneqq\min\left\{
        \bar a,\frac{1}{\sqrt C},
        \sqrt{\frac{\nu E}{4C}}
        \right\},\\
        \kappa_0
        &\coloneqq\max\left\{
        \frac{\rho}{\nu},\frac{4C_0}{\nu E}
        \right\},
    \end{aligned}
    \label{eq:tracking_invariance_parameters}
\end{equation}
where $\rho$ is the prescribed decay rate in
\eqref{eq:averaged_tracking_SPAS}.
Fix $\kappa\geq\kappa_0$ and $0<a\leq a_0$.
By Assumption~\ref{assum:Internal_dynamics},
\eqref{eq:tracking_cW_choice}, and
\eqref{eq:tracking_invariance_parameters}, on the
boundary $W(\bm{\eta}^{av})=c_W$ of $\mathcal S$,
\[
    \begin{aligned}
        \frac{d}{dt}W(\bm{\eta}^{av})
        &\leq-\alpha(c_W)+\sigma(Y)+c_0+Ca^2\\
        &\leq-1.
    \end{aligned}
\]
Similarly, by \eqref{eq:tracking_L_choice} and
\eqref{eq:tracking_invariance_parameters}, on
$|w^{av}|=L$,
\[
    D^+|w^{av}|
    \leq-L+B_{\max}+Ca^2
    \leq-1.
\]
On $|e_\xi^{av}|=E$,
\eqref{eq:tracking_error_inequality} and
\eqref{eq:tracking_invariance_parameters} give,
for almost every $t$,
\[
    \begin{aligned}
        D^+|e_\xi^{av}|
        &\leq-\nu\kappa E+C_0+C\kappa a^2\\
        &\leq-\nu\kappa E
        +\frac{\nu\kappa E}{4}
        +\frac{\nu\kappa E}{4}\\
        &=-\frac{\nu\kappa E}{2}<0.
    \end{aligned}
\]
A first-exit argument using these strict inward
boundary estimates establishes forward invariance
of $\mathcal S$.
Since
\[
    \begin{aligned}
        \bm{\xi}^{av}
        &=\bm{\Theta}^{-1}(\bm{\eta}^{av},y_\xi^{av}),\\
        \hat v^{av}
        &=w^{av}+H(\bm{\eta}^{av},y_\xi^{av}),
    \end{aligned}
\]
the averaged state $(\bm{\xi}^{av},\hat v^{av})$
also remains in a compact set contained in the
domain of definition of the averaged system.
The solution therefore extends to all $t\geq t_0$.

Integrating \eqref{eq:tracking_error_inequality} yields
\begin{equation}
    \begin{aligned}
        |e_\xi^{av}(t)|
        \leq{}&
        e^{-\nu\kappa(t-t_0)}|e_\xi^{av}(t_0)|\\
        &+\frac{C_0+C\kappa a^2}{\nu\kappa}
        \bigl(1-e^{-\nu\kappa(t-t_0)}\bigr)\\
        \leq{}&
        e^{-\nu\kappa(t-t_0)}|e_\xi^{av}(t_0)|\\
        &+\frac{C_0}{\nu\kappa}+\frac{C}{\nu}a^2.
    \end{aligned}
    \label{eq:averaged_tracking_explicit_bound}
\end{equation}

For the prescribed $\delta^{av}>0$, choose
\begin{equation}
    \begin{aligned}
        \kappa^*
        &\coloneqq\max\left\{
        \kappa_0,\frac{2C_0}{\nu\delta^{av}}
        \right\},\\
        a^*
        &\coloneqq\min\left\{
        a_0,\sqrt{\frac{\nu\delta^{av}}{2C}}
        \right\}.
    \end{aligned}
    \label{eq:tracking_final_parameter_choices}
\end{equation}
By \eqref{eq:tracking_invariance_parameters} and
\eqref{eq:tracking_final_parameter_choices}, for all
$\kappa\geq\kappa^*$ and $0<a\leq a^*$,
\[
    \nu\kappa\geq\rho,
    \qquad
    \frac{C_0}{\nu\kappa}+\frac{C}{\nu}a^2
    \leq\delta^{av}.
\]
Substituting these inequalities into
\eqref{eq:averaged_tracking_explicit_bound} gives
\[
    \begin{aligned}
        |e_\xi^{av}(t)|
        &\leq e^{-\rho(t-t_0)}|e_\xi^{av}(t_0)|
        +\delta^{av},\\
        &\hspace{3cm}t\geq t_0.
    \end{aligned}
\]
This proves the lemma.
\hfill$\blacksquare$

\section{Proof of Lemma~\ref{lem:SPES_of_e_xi}}
\label{Proof of Lemma e_xi}
Fix $\Delta_\xi,\rho_\xi,\delta_\xi>0$ and use the construction in the proof of Lemma~\ref{lem:averaged_tracking_SPAS} with $\Delta^{av}=\Delta_\xi$. In particular, let $\mathcal S$ be the compact set in \eqref{eq:tracking_S}. For the transformed system, introduce
\[
    \bm{\eta}=\bm{\vartheta}(\bm{\xi}),
    \qquad
    w=\hat v-H\bigl(\bm{\eta},h(\bm{\xi})\bigr),
\]
where $H$ is defined in \eqref{eq:tracking_H}. The construction of $\mathcal S$ ensures that $(\bm{\eta}(t_0),w(t_0),e_\xi(t_0))\in\mathcal S$ for every admissible initial condition.

Choose
\begin{equation}
    T=\frac{\log 2}{\rho_\xi},
    \qquad
    \delta^{av}=\min\left\{\frac{\delta_\xi}{6},\frac E4\right\}.
    \label{eq:prop_T_delta}
\end{equation}
Choose $\kappa_\xi^*$ and $a_\xi^*$ according to \eqref{eq:tracking_final_parameter_choices}, with $\rho=\rho_\xi$ and this choice of $\delta^{av}$. Reduce $a_\xi^*$, if necessary, so that the flow $\bm{\varphi}(s,\bm{\xi})$ is defined for $|s|\leq a_\xi^*$ and $\bm{\xi}$ in a fixed compact neighbourhood of $\bm{\Theta}^{-1}(\mathcal P)$. Fix any $\kappa\geq\kappa_\xi^*$ and $0<a\leq a_\xi^*$ for the remainder of the proof.

The proof of Lemma~\ref{lem:averaged_tracking_SPAS} establishes forward invariance of $\mathcal S$ for the averaged system. Moreover, \eqref{eq:averaged_tracking_explicit_bound} gives, for every averaged trajectory initialized in $\mathcal S$ at any time $\tau\geq0$,
\begin{equation}
    |e_\xi^{av}(\tau+s)|
    \leq e^{-\rho_\xi s}|e_\xi^{av}(\tau)|+\delta^{av},
    \qquad s\geq0.
    \label{eq:prop_averaged_decay}
\end{equation}

We first establish positive margins from the boundary of $\mathcal S$ at time $\tau+T$. By \eqref{eq:prop_T_delta} and \eqref{eq:prop_averaged_decay},
\begin{equation}
    |e_\xi^{av}(\tau+T)|\leq\frac E2+\delta^{av}
    \leq\frac{3E}{4}.
    \label{eq:prop_error_margin}
\end{equation}
The estimates in the proof of the lemma also give
\begin{equation}
    \begin{aligned}
        \frac{d}{dt}W(\bm{\eta}^{av})
        &\leq-\alpha(W(\bm{\eta}^{av}))
        +\sigma(Y)+c_0+1,\\
        D^+|w^{av}|
        &\leq-|w^{av}|+B_{\max}+1,
    \end{aligned}
    \label{eq:prop_auxiliary_bounds}
\end{equation}
where we used $Ca^2\leq1$. By \eqref{eq:tracking_cW_choice} and continuity of $\alpha$, choose $d_W\in(0,c_W)$ such that
\[
    \alpha(c_W-d_W)\geq\sigma(Y)+c_0+\frac32.
\]
Thus $\frac{d}{dt}W(\bm{\eta}^{av})\leq-1/2$ whenever $c_W-d_W\leq W(\bm{\eta}^{av})\leq c_W$. A trajectory starting below this strip cannot enter it from below. Consequently,
\begin{equation}
    W(\bm{\eta}^{av}(\tau+T))\leq c_W-m_W,
    \label{eq:prop_eta_margin}
\end{equation}
where $m_W=\min\{d_W,T/2\}>0$. Similarly, scalar comparison in \eqref{eq:prop_auxiliary_bounds} yields
\begin{equation}
    |w^{av}(\tau+T)|\leq L-m_w,
    \label{eq:prop_w_margin}
\end{equation}
where $m_w=(L-B_{\max}-1)(1-e^{-T})>0$ by \eqref{eq:tracking_L_choice}. All three margins are uniform in $\tau$ and the averaged initial state in $\mathcal S$.

For any $\tau\geq0$ with $(\bm{\eta}(\tau),w(\tau),e_\xi(\tau))\in\mathcal S$, initialize the averaged trajectory by
\begin{equation}
    \bm{\xi}^{av}(\tau)=\bm{\xi}(\tau),
    \qquad
    \hat v^{av}(\tau)=\hat v(\tau),
    \label{eq:prop_matching_initial_states}
\end{equation}

Invariance of $\mathcal S$ and boundedness of $y_d$ confine the averaged trajectories to a common compact set. For fixed $\kappa,a$, finite-time averaging and the continuously differentiable coordinate transformation therefore ensure that, for every $\varepsilon>0$ and sufficiently large $\omega$, the actual solution exists on $[\tau,\tau+T]$ and
\begin{equation}
    \sup_{0\leq s\leq T}
    \left|
    \begin{pmatrix}
        \bm{\eta}(\tau+s)-\bm{\eta}^{av}(\tau+s)\\
        w(\tau+s)-w^{av}(\tau+s)\\
        e_\xi(\tau+s)-e_\xi^{av}(\tau+s)
    \end{pmatrix}
    \right|
    \leq\varepsilon,
    \label{eq:prop_coordinate_closeness}
\end{equation}
uniformly in $\tau$ and the matching initial state in $\mathcal S$.

Choose
\begin{equation}
    0<\varepsilon\leq
    \min\left\{
        \frac{\delta_\xi}{6},\frac E8,\frac{m_w}{2}
    \right\}.
    \label{eq:prop_epsilon_choice}
\end{equation}
By uniform continuity of $W$ on the relevant compact set, choose $\varepsilon$ smaller if necessary so that
\[
    |\bm{\eta}-\bm{\eta}^{av}|\leq\varepsilon
    \quad\Longrightarrow\quad
    |W(\bm{\eta})-W(\bm{\eta}^{av})|\leq\frac{m_W}{2}.
\]
Choose $\omega^*>0$ so that \eqref{eq:prop_coordinate_closeness} holds with this $\varepsilon$ for every $\omega\geq\omega^*$. For any such $\omega$, combining the averaging estimate with \eqref{eq:prop_error_margin}, \eqref{eq:prop_eta_margin}, and \eqref{eq:prop_w_margin} gives
\begin{equation}
    \begin{aligned}
        |e_\xi(\tau+T)|&\leq\frac{7E}{8}<E,\\
        W(\bm{\eta}(\tau+T))&\leq c_W-\frac{m_W}{2}<c_W,\\
        |w(\tau+T)|&\leq L-\frac{m_w}{2}<L.
    \end{aligned}
    \label{eq:prop_return_to_S}
\end{equation}
Thus every actual trajectory starting in $\mathcal S$ returns to $\mathcal S$ after time $T$ and remains in a common compact neighbourhood between these times.

Let $t_k=t_0+kT$. At each $t_k$, initialize a new averaged trajectory from the actual state as in \eqref{eq:prop_matching_initial_states}. By \eqref{eq:prop_return_to_S}, the same comparison applies on every interval $[t_k,t_{k+1}]$, and the compact bounds establish forward completeness. For $0\leq s\leq T$, equations \eqref{eq:prop_averaged_decay} and \eqref{eq:prop_coordinate_closeness} imply
\begin{equation}
    \begin{aligned}
        |e_\xi(t_k+s)|
        &\leq |e_\xi^{av}(t_k+s)|+\varepsilon\\
        &\leq e^{-\rho_\xi s}|e_\xi(t_k)|
        +\delta^{av}+\varepsilon.
    \end{aligned}
    \label{eq:prop_interval_error}
\end{equation}
Taking $s=T$ and using $e^{-\rho_\xi T}=1/2$ gives
\[
    |e_\xi(t_{k+1})|
    \leq\frac12|e_\xi(t_k)|+\delta^{av}+\varepsilon.
\]
Iteration yields, for every $k\geq0$,
\begin{equation}
    |e_\xi(t_k)|
    \leq2^{-k}|e_\xi(t_0)|+2(\delta^{av}+\varepsilon).
    \label{eq:prop_sampled_error}
\end{equation}

Finally, write any $t\geq t_0$ as $t=t_k+s$ with $0\leq s<T$. Since $2^{-k}=e^{-\rho_\xi kT}$, \eqref{eq:prop_interval_error} and \eqref{eq:prop_sampled_error} give
\[
    \begin{aligned}
        |e_\xi(t)|
        &\leq e^{-\rho_\xi(t-t_0)}|e_\xi(t_0)|\\
        &\quad+(2e^{-\rho_\xi s}+1)(\delta^{av}+\varepsilon)\\
        &\leq e^{-\rho_\xi(t-t_0)}|e_\xi(t_0)|+\delta_\xi,
    \end{aligned}
\]
where the last inequality follows from \eqref{eq:prop_T_delta} and \eqref{eq:prop_epsilon_choice}. This proves \eqref{eq:shifted_system_SPAS}.
\hfill$\blacksquare$

% \bibliographystyle{plain}        % Include this if you use bibtex 
% \bibliography{autosam}           % and a bib file to produce the 

\end{document}